\documentclass[english]{paper}
\usepackage[T1]{fontenc}
\usepackage[utf8x]{inputenc}
\usepackage[letterpaper]{geometry}
\usepackage{beramono}
\usepackage{fancyhdr}
\usepackage{color}
\usepackage{babel}
\usepackage{array}
\usepackage{longtable}
\usepackage{float}
\usepackage{units}
\usepackage{textcomp}
\usepackage{multirow}
\usepackage{amsmath}
\usepackage{fixltx2e}
\usepackage{graphicx}
\PassOptionsToPackage{version=3}{mhchem}
\usepackage{mhchem}
\usepackage[unicode=true,
 bookmarks=true,bookmarksnumbered=false,bookmarksopen=false,
 breaklinks=true,pdfborder={0 0 0},backref=false,colorlinks=false]
 {hyperref}
\hypersetup{
 pdfauthor={Wendell Smith}}

\makeatletter

\usepackage{savefnmark}
\usepackage{authblk}
\usepackage{multicol}
\usepackage{caption}
\usepackage{subcaption}

\begin{document}
\renewcommand\Affilfont{\itshape\small}

\author[1]{W. Wendell Smith}
\author[1]{Carl F. Schreck}
\author[1]{Nabeem Hashem}
\author[1]{Sherwin Soltani}
\author[2]{Abhinav Nath}
\author[2,1]{Elizabeth Rhoades}
\author[3,1]{Corey S. O'Hern}

\affil[1]{Department of Physics, Yale University, New Haven, CT}
\affil[2]{Department of Molecular Biophysics and Biochemistry, Yale University, New Haven, CT}
\affil[3]{Department of Mechanical Engineering and Materials Science, Yale University, New Haven, CT}

\title{Molecular Simulations of the Fluctuating Conformational
Dynamics of Intrinsically Disordered Proteins} \maketitle
\begin{abstract}
Intrinsically disordered proteins (IDPs) do not possess well-defined
three-dimensional structures in solution under physiological
conditions.  We develop all-atom, united-atom, and coarse-grained
Langevin dynamics simulations for the IDP $\alpha$-synuclein that
include geometric, attractive hydrophobic, and screened electrostatic
interactions and are calibrated to the inter-residue separations
measured in recent smFRET experiments.  We find that
$\alpha$-synuclein is disordered with conformational statistics that
are intermediate between random walk and collapsed globule behavior.
An advantage of {\it calibrated} molecular simulations over constraint
methods is that physical forces act on all residues, not only on
residue pairs that are monitored experimentally, and these simulations
can be used to study oligomerization and aggregation of multiple
$\alpha$-synuclein proteins that may precede amyloid formation.
\end{abstract}
\begin{multicols}{2}

\section{Introduction}

Intrinsically disordered proteins (IDPs) do not possess well-defined
three-dimensional structures in physiological conditions. Instead,
IDPs can range from collapsed globules to extended chains with highly
fluctuating conformations in aqueous
solution~\cite{vucetic_flavors_2003}.  IDPs play a significant role in
cellular signaling and control since they can interact with a wide
variety of binding targets~\cite{sugase_mechanism_2007}.  In addition,
their propensity to aggregate to form oligomers and fibers has been
linked to the onset of amyloid
diseases~\cite{uversky_intrinsically_2008}.  The conformational and
dynamic heterogeneity of IDPs makes their structural characterization
by traditional biophysical approaches challenging.  Also, force fields
employed in all-atom molecular dynamics simulations, which are
typically calibrated for folded proteins, can yield results that
differ significantly from experiments~\cite{dedmon_mapping_2004}.

In this manuscript, we focus on the IDP $\alpha$-synuclein, which is a
$140$-residue neuronal protein linked to Parkinson's disease and Lewy
body dimentia~\cite{vilar_fold_2008}. Previous NMR studies have found
that $\alpha$-synuclein is largely unfolded in solution, but more
compact than a random coil with same
length~\cite{dedmon_mapping_2004,eliezer_conformational_2001,li_conformational_2002}.
The precise mechanism for aggregation in $\alpha$-synuclein has not
been identified, although it is known that aggregation is enhanced at
low
pH~\cite{tsigelny_dynamics_2007,li_conformational_2002,uversky_effects_2005},
possibly due to the loss of long-range contacts between the N- and C-
termini of the protein~\cite{ullman_explaining_2011}.

Quantitative structural information has been obtained for
$\alpha$-synuclein using single-molecule fluorescence resonance energy
transfer (smFRET) between more than twelve donor and acceptor
pairs~\cite{trexler_single_2010}.  These experimental studies have
measured inter-residue separations for both the neutral and low pH
ensembles. Prior studies have implemented the inter-residue
separations from smFRET as constraints in Monte Carlo simulations with
only geometric ({\it e.g.} bond-length and bond-angle) and repulsive
Lennard-Jones interactions to investigate the natively disordered
ensemble of conformations for monomeric
$\alpha$-synuclein~\cite{Nath_conformational_2012}. In contrast, we
develop all-atom, united-atom, and coarse-grained Langevin dynamics
simulations of $\alpha$-synuclein that include geometric, attractive
hydrophobic, and screened electrostatic interactions.  The simulations
are calibrated to closely match the inter-residue separations from the
smFRET experiments.  An advantage of this method over constrained
simulations is that physical forces, which act on all 
residues in the protein, are tuned so that the inter-residue
separations from experiments and simulations agree.  In future
studies, we will employ these calibrated Langevin dynamics simulations
to study oligomerization and aggregation of multiple
$\alpha$-synuclein proteins over a range of solvent conditions.

\begin{figure*}
\centerline{\begin{minipage}{.95\textwidth}
\small{\texttt{\textbf{\textcolor{blue}{MET~ASP~VAL~PHE~MET~LYS~GLY~LEU~SER~LYS~ALA~LYS~GLU~GLY~VAL~VAL~ALA~ALA~ALA~GLU}}~20~\\
\textbf{\textcolor{blue}{LYS~THR~LYS~GLN~GLY~VAL~ALA~GLU~ALA~ALA~GLY~LYS~THR~LYS~GLU~GLY~VAL~LEU~TYR~VAL}}~40~\\
\textbf{\textcolor{blue}{GLY~SER~LYS~THR~LYS~GLU~GLY~VAL~VAL~HIS~GLY~VAL~ALA~THR~VAL~ALA~GLU~LYS~THR~LYS}}~60~\\
GLU~GLN~VAL~THR~ASN~VAL~GLY~GLY~ALA~VAL~VAL~THR~GLY~VAL~THR~ALA~VAL~ALA~GLN~LYS~80~\\
THR~VAL~GLU~GLY~ALA~GLY~SER~ILE~ALA~ALA~ALA~THR~GLY~PHE~VAL~\textit{\textcolor{red}{LYS~LYS~ASP~GLN~LEU}}~100\\
\textit{\textcolor{red}{GLY~LYS~ASN~GLU~GLU~GLY~ALA~PRO~GLN~GLU~GLY~ILE~LEU~GLU~ASP~MET~PRO~VAL~ASP~PRO}}~120\\
\textit{\textcolor{red}{ASP~ASN~GLU~ALA~TYR~GLU~MET~PRO~SER~GLU~GLU~GLY~TYR~GLN~ASP~TYR~GLU~PRO~GLU~ALA}}~140}}\\
\end{minipage}}
\caption{\label{regions}The three main regions of the $140$-residue
protein $\alpha$-synuclein. Residues $1$-$60$ form the highly basic
N-terminal region (bold, blue), residues $61$-$95$ form the
hydrophobic central region (plain text), and residues $96$-$140$ form
the acidic C-terminal region (italics,
red)~\cite{ullman_explaining_2011,trexler_single_2010}.}
\end{figure*}

\section{Methods}
\label{method}

The $140$-residue IDP $\alpha$-synuclein includes a negatively charged
N-terminal region, hydrophobic central region, and positively charged
C-terminal region (Fig.~\ref{regions}) at neutral pH.  We study three
models for $\alpha$-synuclein with different levels of geometric
complexity: a) all-atom, b) united-atom, and c) coarse-grained, as
shown in Fig.~$2$.

\begin{figure*}
\label{fig:CG-vis}
\hfill
\begin{subfigure}[b]{2in}
\centering
\includegraphics[height=3in,keepaspectratio]{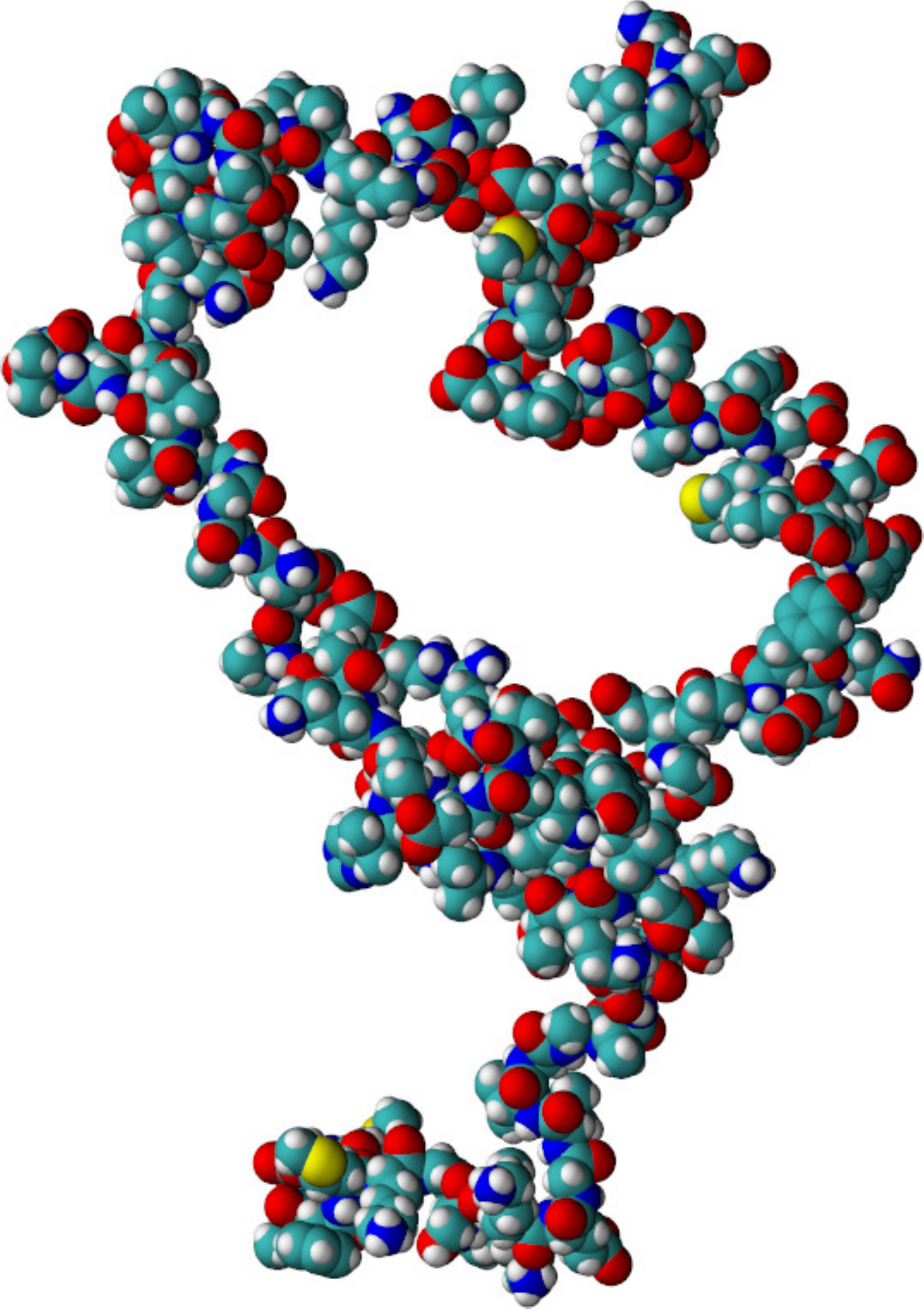}
\end{subfigure}
\hfill
\begin{subfigure}[b]{2in}
\centering
\includegraphics[height=3in,keepaspectratio]{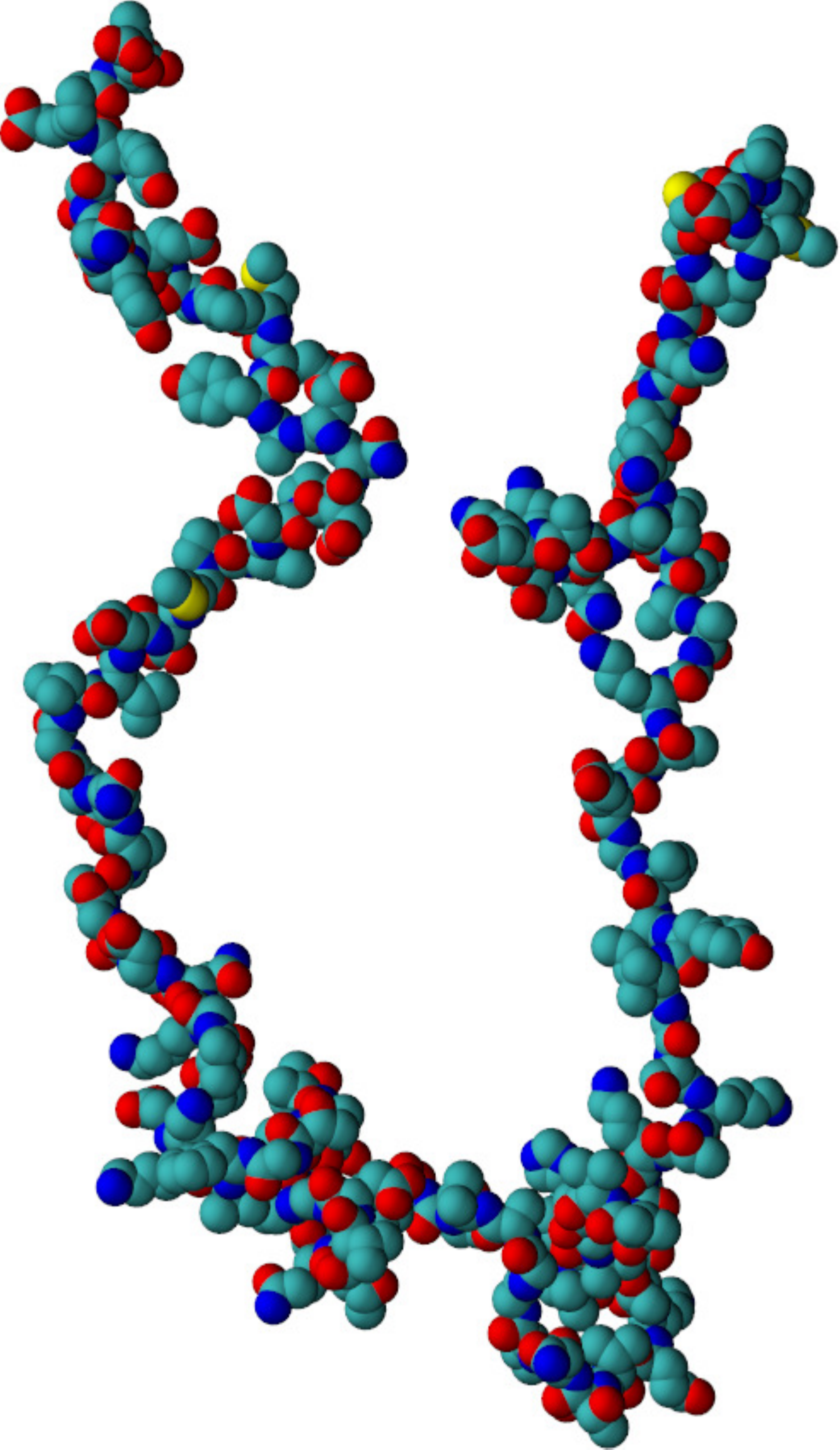}
\end{subfigure}
\hfill
\begin{subfigure}[b]{2in}
\centering
\includegraphics[height=3in,keepaspectratio]{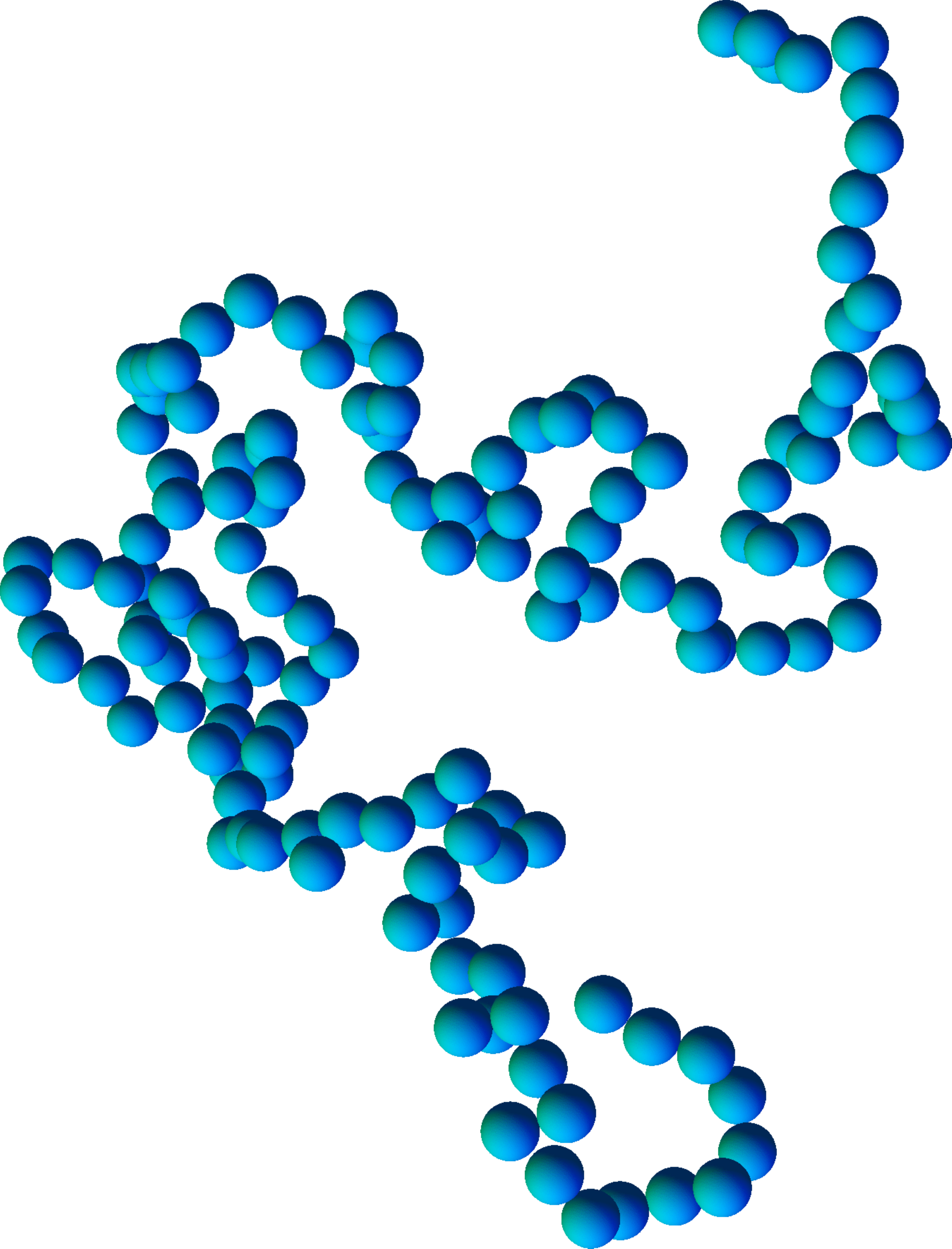}
\end{subfigure}
\hfill

\caption{Snapshots of the (left) all-atom, (center) united-atom, and
(right) coarse-grained representations of $\alpha$-synuclein from
Langevin dynamics simulations at temperature $T_0 = 293 {\rm K}$, pH $7.4$,
and ratio of hydrophobic to electrostatic interactions $\alpha =
1.1$. For the all-atom and united-atom models, hydrogen, carbon,
oxygen, nitrogen, and sulfur atoms are colored white, cyan, red, blue,
and yellow, respectively.  For the coarse-grained model,
each blue-shaded monomer represents an amino acid.}
\end{figure*}

\subsection{All-Atom Model}

The all-atom model (including hydrogen atoms) matches closely the
geometric properties of proteins.  The average bond lengths
$\langle l_{ij} \rangle$, bond angles $\langle \theta_{ijk}\rangle$,
and backbone dihedral angle $\omega$ between atoms
$C_{\alpha}$-$C$-$N$-$C_{\alpha}$ on successive residues were obtained
from the Dunbrack database of $850$ high-resolution protein crystal
structures~\cite{dunbrack_jr._bayesian_1997}.  The
$242$ distinct bonds and $440$ distinct bond angles in
$\alpha$-synuclein were fixed using the following spring potentials:
\begin{equation}
\label{bond-length}
V^{bl} = \frac{k_l}{2} \sum_{ij} (r_{ij} - \langle l_{ij} \rangle)^2,
\end{equation}
where $k_l$ is the bond-length stiffness and $r_{ij}$ is the
center-to-center separation between bonded atoms $i$ and $j$, and 
\begin{equation}
\label{bond-angle}
V^{ba} = \frac{k_{\theta}}{2} \sum_{ijk} (\theta_{ijk} - \langle \theta_{ijk}
\rangle)^2,
\end{equation}
where $k_{\theta}$ is the bond-angle stiffness and $\theta_{ijk}$ is
the angle between bonded atoms $i$, $j$, and $k$.  The average
backbone dihedral angle between the
$C_{\alpha}$-$C$-$N$-$C_{\alpha}$ atoms was constrained to
zero using
\begin{equation}
\label{dihedral_angle}
V^{da} = \frac{k_{\omega}}{2} \sum_{ijkl} \omega_{ijkl}^2. 
\end{equation}
We chose $k_l= 5 \times 10^3~k_b T_0/\mbox{Å}^2$ and
$k_\theta=k_{\omega} = 2 \times 10^5~k_b T_0/{\rm rad}^2$ (with $T_0 =
293 {\rm K}$) so that the root-mean-square (rms) fluctuations in the
bond lengths, bond angles, and dihedral angles were below $0.05
~\mbox{Å}$ and $0.008~{\rm rad}$, respectively. These rms values
occur in the protein crystal structures from the Dunbrack database.
Note that no explicit interaction potentials were used to constrain
the backbone dihedral angles $\phi$ and $\psi$ and side-chain dihedral
angles. However, the bond lengths, bond angles, and sizes of the atoms
were were calibrated so that they take on physical values. (See
Appendix~\ref{apx:atom-sizes}.)

\begin{figure*}
\hfill
\begin{subfigure}[b]{2in}
\centering
\includegraphics[width=\textwidth,keepaspectratio]{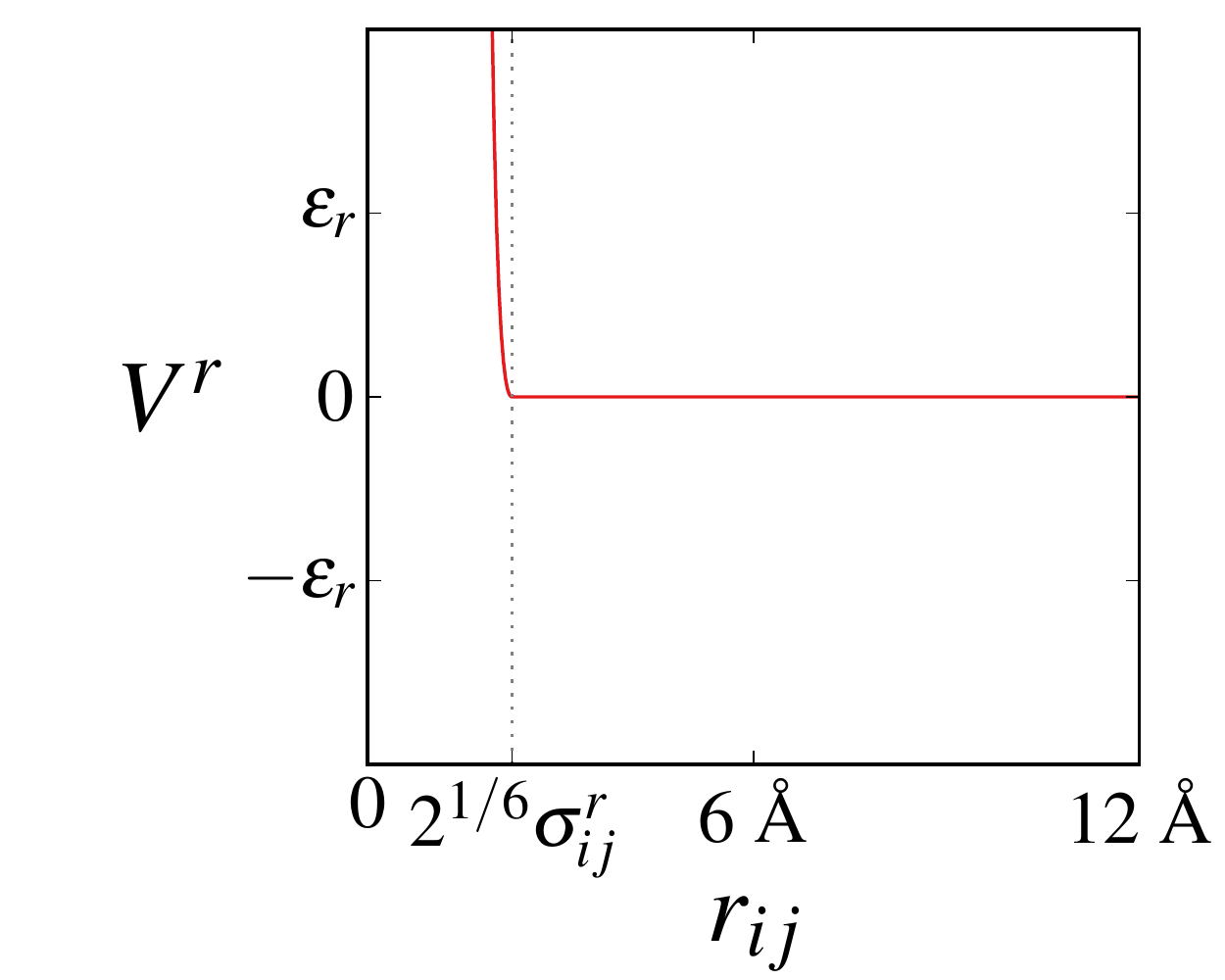}
\end{subfigure}
\hfill
\begin{subfigure}[b]{2in}
\centering
\includegraphics[width=\textwidth,keepaspectratio]{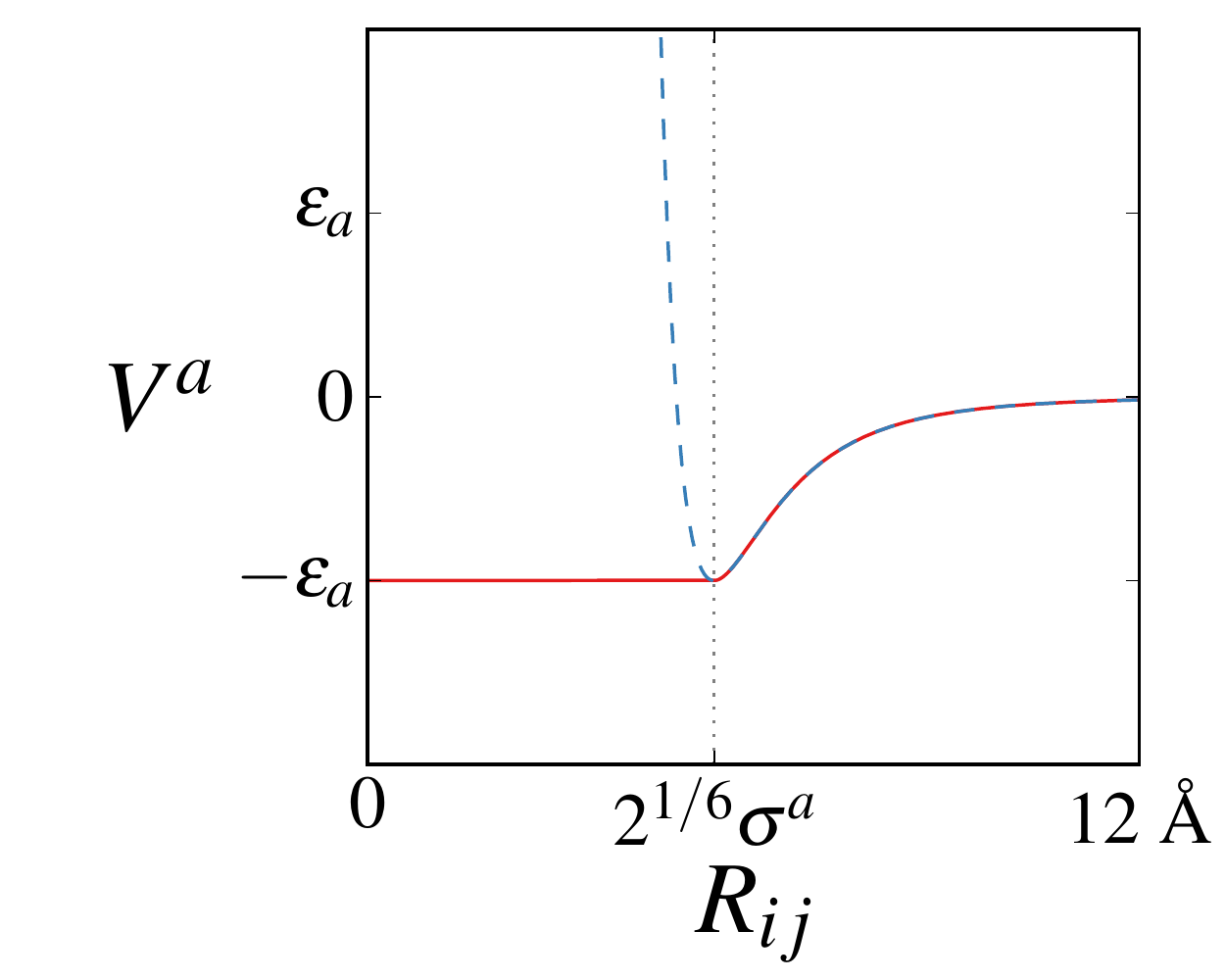}
\end{subfigure}
\hfill
\begin{subfigure}[b]{2in}
\centering
\includegraphics[width=\textwidth,keepaspectratio]{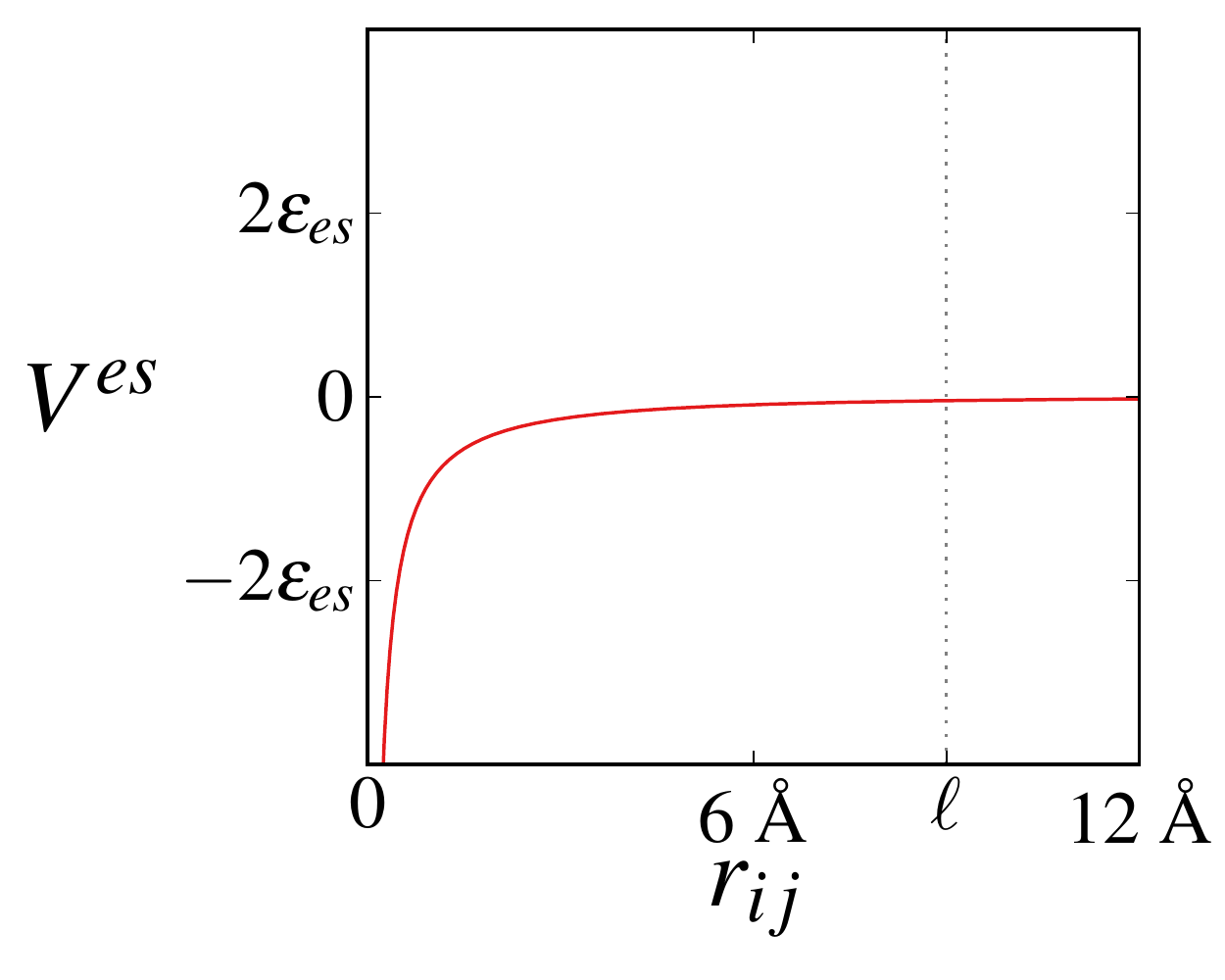}
\end{subfigure}
\hfill
\caption{\label{fig:3}Schematics of (a) the purely repulsive
Lennard-Jones potential $V^r$ in Eq.~\ref{repulsive} (solid line),
(b) attractive Lennard-Jones potential $V^a$ in Eq.~\ref{attractive}
(solid line), and (c) screened Coulomb potential $V^{es}$ in
Eq.~\ref{screened} (solid line).  The dashed line in (b) represents
repulsive Lennard-Jones interactions between residues $i$ and $j$
in the coarse-grained model.}
\end{figure*}

We included three types of interactions between nonbonded atoms: 1)
the purely repulsive Lennard-Jones potential $V^r$ to model steric
interactions, 2) attractive Lennard-Jones interactions $V^a$ between
$C_{\alpha}$ atoms on each residue to model hydrophobicity, and 3)
screened electrostatic interactions $V^{es}$ between atoms in the
charged residues LYS, ARG, HIS, ASP, and GLU.  Thus, the total
interaction energy is $V=V^{bl} + V^{ba} + V^{da} +
V^{r}+V^{a}+V^{es}$. (See Fig.~\ref{fig:3}.)

The purely repulsive Lennard-Jones potential is 
\begin{multline}
\label{repulsive}
V^{r}=\epsilon_{r}\left(4\left[\left(\frac{\sigma_{ij}^{r}}{r_{ij}}\right)^{12}-\left(\frac{\sigma_{ij}^{r}}{r_{ij}}\right)^{6}\right]+1\right)\\
\times\Theta\left(2^{1/6}\sigma_{ij}^{r}-r_{ij}\right),
\end{multline}
where $\Theta\left(x\right)$ is the Heaviside step function that sets
$V^r=0$ for $r_{ij} \ge 2^{1/6} \sigma^r_{ij}$, $\epsilon_r/k_b T_0
=1$, and $\sigma_{ij}^{r}=(\sigma_{i}^{r}+\sigma_{j}^{r})/2$ is the
average diameter of atoms $i$ and $j$. We used the atom sizes (for
hydrogen, carbon, oxygen, nitrogen, and sulfur) from
Ref.~\cite{zhou_power_2012} after verifying that the backbone dihedral
angles for the all-atom model sample the sterically allowed $\phi$ and
$\psi$ values in the Ramachandran map~\cite{ramachandran_1963} when
$V=V^{bl} + V^{ba} + V^{da} + V^{r}$. (See
Appendix~\ref{apx:atom-sizes}.)

The hydrophobic interactions between residues were modeled
using the attractive Lennard-Jones potential
\begin{multline}
\label{attractive}
V^{a}=\epsilon_{a} \sum_{ij} \left[ \lambda_{ij}\left(4\left[\left(\frac{\sigma^{a}}{R_{ij}}\right)^{12}-\left(\frac{\sigma^{a}}{R_{ij}}\right)^{6}\right]+1\right)\right.\\
\left.\times\Theta\left(R_{ij}-2^{1/6}\sigma^{a}\right) - \lambda_{ij}\right],
\end{multline}
where $\epsilon_{a}$ is the attraction strength, $R_{ij}$ is the 
center-to-center separation between $C_{\alpha}$ atoms on residues $i$ and $j$,
\begin{equation}
\lambda_{ij}=\sqrt{h_{i}h_{j}}\label{eq:hphob-lambda},
\end{equation}
$h_{i}$ is the hydrophobicity index for residue $i$ that ranges from
$0$ (hydrophilic) to $1$ (hydrophobic) in Table~ \ref{table-polarity},
and $\sigma^{a} \approx 4.8~\mbox{Å}$ is the typical separation
between centers of mass of neighboring residues.  We
find that the results for the conformational statistics 
for $\alpha$-synuclein are not sensitive to small changes in $\sigma^{a}$
and $h_{i}$ (Appendix~\ref{apx:robust}).

The screened Coulomb potential was used to model the electrostatic 
interactions between atoms $i$ and $j$ for $\alpha$-synuclein in water: 
\begin{equation}
\label{screened}
V^{es}=\epsilon_{es} \sum_{ij}
\frac{q_{i}q_{j}}{e^2} \frac{\sigma^a}{r_{ij}}e^{-\frac{r_{ij}}{\ell}},
\end{equation}
where $e$ is the fundamental charge, $\epsilon_{es} = 
\sigma^a e^2/4 \pi \epsilon_0 \epsilon$, $\epsilon_0$ is
the vacuum permittivity, $\epsilon=80$ is the permittivity of water,
and $\ell=9~\mbox{Å}$ is the Coulomb screening length in an
aqueous solution with a $150 {\rm mM}$ salt concentration.  The
partial charge $q_i$ on atom $i$ in one of the charged residues 
LYS, ARG, HIS, ASP, and GLU is given in Table~\ref{table-charge}.

\begin{table*}
\noindent \begin{centering}
\begin{tabular}{|c|c|c|c|c|c|c|c|c|c|}
\hline 
ALA & ARG & ASN & ASP & CYS & GLN & GLU & GLY & HIS & ILE\\
\hline 
\hline 
0.735 & 0.37 & 0.295 & 0.41 & 0.76 & 0.41 & 0.54 & 0.5 & 0.29 & 1\\
\hline 
\noalign{\vskip1em}
\hline 
LEU & LYS & MET & PHE & PRO & SER & THR & TRP & TYR & VAL\\
\hline 
\hline 
0.985 & 0.385 & 0.87 & 1 & 0.27 & 0.475 & 0.565 & 0.985 & 0.815 & 0.88\\
\hline 
\end{tabular}
\par\end{centering}

\caption{\label{table-polarity}Hydrophobicity indices $h_{i}$ that range
from $0$ (hydrophilic) to $1$ (hydrophobic) for residues in
$\alpha$-synuclein at pH $7.4$~\cite{monera_relationship_1995}.}
\end{table*}

\noindent \begin{center}
\begin{table*}
\noindent \begin{centering}
\begin{tabular}{|c|c|c|c|}
\hline 
Residue & Atom & Atom Charge $q_i$ & Residue Charge $Q_i$\\
\hline 
LYS & $\ce{N_{\zeta}}$ & 1 & 1\\
\hline 
\multirow{3}{*}{ARG} & $\ce{N_{\eta1}}$ & 0.39 & \multirow{3}{*}{1}\\
\cline{2-3} 
 & $\ce{N_{\eta2}}$ & 0.39 & \\
\cline{2-3} 
 & $\ce{N_{\varepsilon}}$ & 0.22 & \\
\hline 
\multirow{2}{*}{HIS} & $\ce{N^{\delta1}}$ & 0.05 & \multirow{2}{*}{0.1}\\
\cline{2-3} 
 & $\ce{N^{\varepsilon2}}$ & 0.05 & \\
\hline 
\multirow{2}{*}{ASP} & $\ce{O^{\delta1}}$ & -0.5 & \multirow{2}{*}{-1}\\
\cline{2-3} 
 & $\ce{O^{\delta2}}$ & -0.5 & \\
\hline 
\multirow{2}{*}{GLU} & $\ce{O^{\varepsilon1}}$ & -0.5 & \multirow{2}{*}{-1}\\
\cline{2-3} 
 & $\ce{O^{\varepsilon2}}$ & -0.5 & \\
\hline 
\end{tabular}\par\end{centering}

\caption{\label{table-charge}Partial charges $q_i$ on atom $i$ (left)
and total charge $Q_i$ on residue $i$ (right) for the charged residues
LYS, ARG, HIS, ASP, and GLU at pH $7.4$~\cite{oostenbrink_2004}.
The total partial charge $q = \sum_i q_i$ for the N-terminal, central,
and C-terminal regions are $4.1$, $-1$, and $-12.0$, respectively.}
\end{table*}

\par\end{center}

\subsection{United-Atom Model}

For the united-atom model, we do not explicitly model the hydrogen
atoms.  Instead, we use a set of $11$ atom sizes $\sigma_{i}^r$ from
Ref.~\cite{richards_interpretation_1974}, where the hydrogens are
subsumed into the heavy atoms: C ($\sigma_i^r/2=1.53~\mbox{Å}$), CH
($1.80 ~\mbox{Å}$), C${\rm H}_2$ ($1.80 ~\mbox{Å}$), C${\rm H}_3$
($1.80 ~\mbox{Å}$), O ($1.26 ~\mbox{Å}$), OH ($1.44 ~\mbox{Å}$), N
($1.53 ~\mbox{Å}$), N${\rm H}_2$ ($1.57 ~\mbox{Å}$), N${\rm H}_3$
($1.80 ~\mbox{Å}$), and S ($1.62 ~\mbox{Å}$).  We optimized the atom
sizes by characterizing the backbone dihedral angles $\phi$ and $\psi$
as a function of $\sigma_{i}^r$ in the united-atom simulations with
$V=V^{bl} + V^{ba} + V^{da} + V^{r}$.  The $\phi$ and $\psi$ backbone
dihedral angle distributions closely match that from the Ramachandran
map ({\it i.e.} the $\alpha$-helix and $\beta$-sheet regions) when we
scale the atom sizes in Ref.~\cite{richards_interpretation_1974} by
$0.9$ as shown in Appendix~ \ref{apx:atom-sizes}.  Otherwise, the
all-atom and united-atom models use the same interaction potentials in
Eqs.~\ref{bond-length}-\ref{screened}.

\subsection{Coarse-Grained Model}
\label{coarse}

For the coarse-grained model, we employed a backbone-only $C_{\alpha}$
representation of $\alpha$-synuclein where each residue $i$ is
represented by a spherical monomer $i$ with size $\sigma^a$, mass $M$,
hydrophobicity $h_i$, and charge $Q_i$. The average bond length
between monomers $i$ and $j$ was fixed to $\langle l_{ij} \rangle =
4.0 ~\mbox{Å}$, which is the average separation between $C_{\alpha}$ atoms 
on neighboring residues, using Eq.~\ref{bond-length} (with $r_{ij}$ replaced
by $R_{ij}$).  The bond-angle $\Theta$ (between three successive
$C_{\alpha}$ atoms) and dihedral-angle $\Phi$ (between four successive
$C_{\alpha}$ atoms) potentials were calculated so that the $\Theta$
and $\Phi$ distributions matched those from the united-atom
simulations with $V=V^{bl} + V^{ba} + V^{da} + V^{r}$.  The $\Theta$
distributions from the united-atom model were approximately Gaussian
with mean $\langle \Theta \rangle = 2.13~{\rm rad}$ and standard deviation
$\sigma_{\Theta}= 0.345~{\rm rad}$.

The dihedral angle potential $V^{da}$ for the coarse-grained 
simulations was obtained by fitting the distribution 
$P(\Phi)$ from the united-atom simulations to a seventh-order Fourier series
\begin{eqnarray*}
V^{da}(\Phi) & = & \sum_{k=0}^{6}a_{k}\cos\left(k\Phi\right)+b_{k}\sin\left(k\Phi\right),
\end{eqnarray*}
where $a_{k} = -2k_b T_0\left\langle
\cos\left(k\Phi\right) \log
P\left(\Phi\right) \right\rangle$, $b_k = -2k_b T_0 \left\langle
\sin\left(k\Phi\right) \log
P\left(\Phi\right) \right\rangle$, and the angle brackets
indicate an average over time and dihedral angles along the
protein backbone.

For steric interactions between residues, we used the purely repulsive
Lennard-Jones potential in Eq.~\ref{repulsive} with $r_{ij}$ and
$\sigma^r_{ij}$ replaced by $R_{ij}$ and $\sigma^a$ respectively.  The
hydrophobic interactions are the same as those in
Eqs.~\ref{attractive} and~\ref{eq:hphob-lambda} with $\epsilon_a =
\epsilon_r$.  The electrostatic interactions between residues are
given by Eq.~\ref{screened} with $q_i$ and $r_{ij}$ replaced by $Q_i$
and $R_{ij}$, respectively.

\subsection{Langevin Dynamics}
\label{langevin}

The all-atom, united-atom, and coarse-grained models were simulated at
fixed $NVT$ using a Langevin thermostat~\cite{ermak}, modified
velocity Verlet integration scheme, and free boundary conditions. We
set the time step $\Delta t = 10^{-2} t_0$ and damping coefficient
$\gamma= 10^{-3} t_0^{-1}$, where $t_0 = \sqrt{m \langle \sigma_{ij}^r
\rangle/\epsilon_r}$ and $m$ is the hydrogen mass for the all-atom and
united-atom models and $t_0 = \sqrt{M \sigma^a/\epsilon_r}$ for the
coarse-grained model.  The initial atomic positions were obtained from
a micelle-bound NMR structure (protein data bank identifier 1XQ8) for
$\alpha$-synuclein at pH $7.4$ and temperature $298 {\rm
K}$~\cite{ulmer_structure_2005}.  The initial positions for the
coarse-grained model were obtained from simulations at high
temperature with only bond-length, bond-angle, and dihedral-angle
constraints and repulsive Lennard-Jones interactions.  The simulations
were run for times much longer than the characteristic relaxation time
from the decay of the radius of gyration autocorrelation function.

In the results below, we will study the radius of gyration $R_g$ and
distribution of inter-residue separations $P(R_{ij})$ as a function of
the ratio of the attractive hydrophobic and electrostatic energy
scales $\alpha = \epsilon_a/\epsilon_{es}$ and quantitatively compare
the results from smFRET experiments and all-atom, united-atom, and
coarse-grained simulations.

\section{Results}
\label{results}

\begin{figure*}
\noindent \begin{centering}
\includegraphics[width=2.5in]{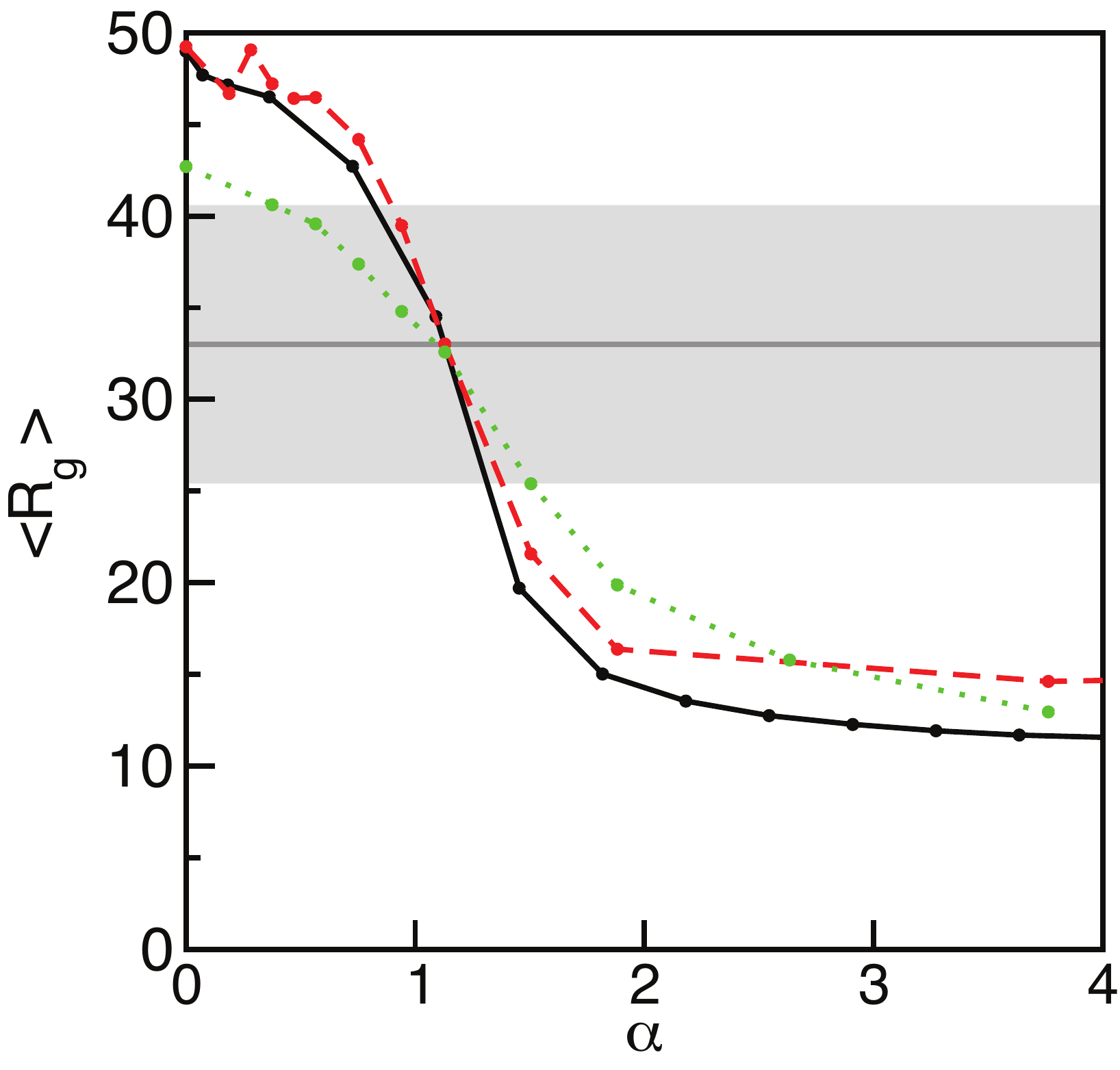}
\par\end{centering}
\caption{\label{Rg-vs-alpha}Average radius of gyration $\langle R_{g}
\rangle$ versus the ratio of the attractive hydrophobic to
electrostatic interactions $\alpha$ for the coarse-grained (black
solid), united-atom (red dashed), and all-atom (green dotted) models
at $T_0$ (or the temperature that gives $R_g \approx 33~\mbox{Å}$
in the coarse-grained simulations) and pH $7.4$. The horizontal line
and gray shaded region indicate the average and standard deviation
over recent NMR, SAXS, and smFRET experimental measurements, $\langle
R_g \rangle = 33.0 \pm 7.7 ~\mbox{Å}$, for monomeric
$\alpha$-synuclein near $T_0$ and neutral
pH~\cite{dedmon_mapping_2004,li_conformational_2002,uversky_effects_2005,Nath_conformational_2012,uversky_trimethylamine-n-oxide-induced_2001,tashiro_characterization_2008,rekas_structure_2010,salmon_nmr_2010}.}
\end{figure*}

\begin{figure*}
\centering
\includegraphics[width=\textwidth]{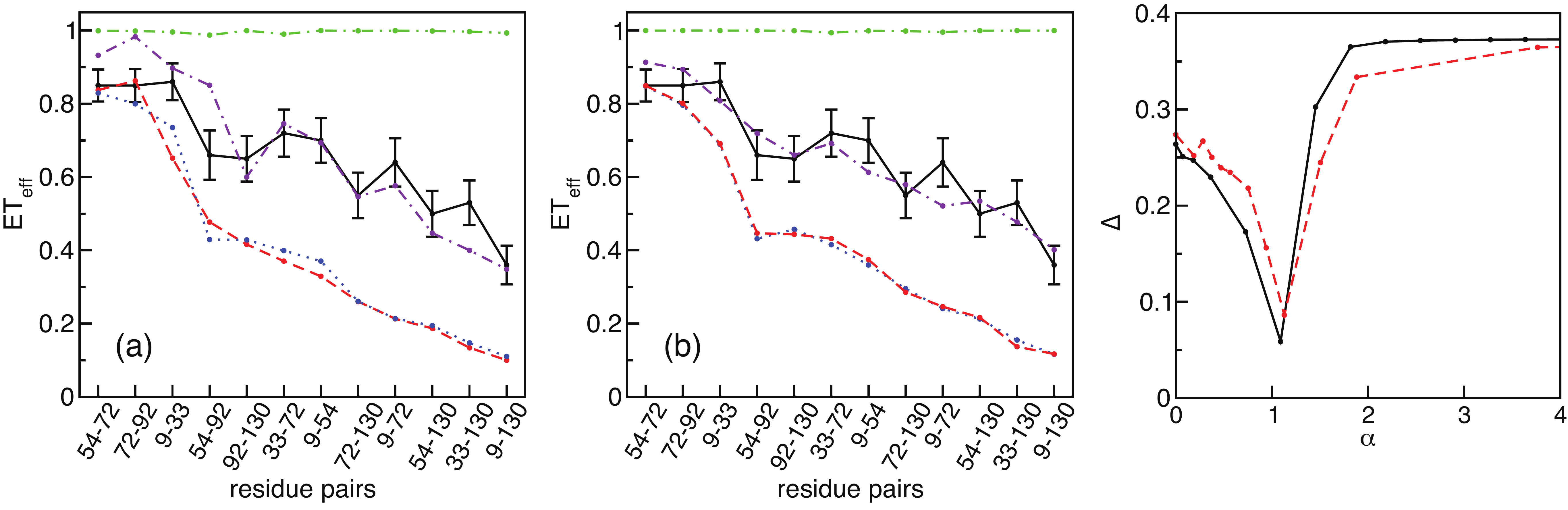}

\caption{\label{fig-ETeffs}A comparison of FRET efficiencies
$ET_{\textrm{eff}}$ for twelve residue pairs from simulations and
experiments of $\alpha$-synuclein.  In (a) the data includes FRET
efficiencies from united-atom simulations of a random walk (red
dashed), collapsed globule (green dot-dot-dashed), only electrostatic
interactions at temperature $T_0$ (blue dotted), and ratio of
attractive hydrophobic to electrostatic interactions $\alpha=1.1$
(purple dot-dashed) at $T_0$ and recent smFRET
experiments~\cite{Nath_conformational_2012} (black solid). Error bars
from experiments were calculated using a resampling method that
accounts for uncertainty in the determination of $ET_{\rm eff}$
($1$-$2\%$) and variations in $R_0$ ($7$-$8\%$) due to the effects
of the protein environment on the smFRET fluorophores. In (b) we
compare the FRET efficiencies from recent smFRET
experiments~\cite{Nath_conformational_2012} to the coarse-grained
simulations of a random walk (red dashed), collapsed globule (green
dot-dot-dashed), only electrostatics interactions (blue dotted), and
both attractive hydrophobic and electrostatic interactions with
$\alpha=1.1$ at a temperature that yields $\langle R_g \rangle \approx
33 ~\mbox{Å}$ (purple dot-dashed). (c) The rms deviation $\Delta$
between the FRET efficiencies from the united-atom simulations and
smFRET experiments (black solid) and the coarse-grained simulations
and smFRET experiments (red dashed) versus $\alpha$.  The minimum rms
$\Delta_{\rm min} \approx 0.09$ occurs near $\alpha \approx 1.1$ for
both the united-atom and coarse-grained simulations.}
\end{figure*}

In Fig.~\ref{Rg-vs-alpha}, we show the radius of gyration that
characterizes the overall protein shape for the all-atom, united-atom,
and coarse-grained models,
\begin{equation}
R_g = \sqrt{\frac{1}{N} \sum_{i=1}^{N} ({\vec r}_i - \langle {\vec r}_i\rangle)^2  },
\end{equation}
where ${\vec r}_i$ is the position of atom or monomer $i$, as a
function of the ratio of the attractive hydrophobic to electrostatic
interactions $\alpha$ at temperature $T_0$ and pH $7.4$.  For $\alpha
\gg 1$, the protein forms a collapsed globule with $\langle R_g
\rangle \approx 12$-$15 ~\mbox{Å}$.  Whereas for $\alpha \ll 1$, the
models only include electrostatics interactions, and $\langle R_g
\rangle$ is similar to the random walk values for the three models
(all-atom: $42.8 ~\mbox{Å}$, united-atom: $48.6 ~\mbox{Å}$,
coarse-grained: $48.2~\mbox{Å}$).  The crossover between random
walk and collapsed globule behavior for $\langle R_{g} \rangle$ occurs
near $\alpha \approx 1$.

A number of recent SAXS, NMR, and smFRET experiments have measured the
radius of gyration for monomeric $\alpha$-synuclein near $T_0$ and
neutral
pH~\cite{dedmon_mapping_2004,li_conformational_2002,uversky_effects_2005,Nath_conformational_2012,uversky_trimethylamine-n-oxide-induced_2001,tashiro_characterization_2008,rekas_structure_2010,salmon_nmr_2010}.
As shown in Fig.~\ref{Rg-vs-alpha}, the average over these
experimental measurements is $\langle R_g \rangle = 33.0 \pm 7.7
~\mbox{Å}$, and thus the $\langle R_g \rangle$ for
$\alpha$-synuclein falls in between the random walk and collapsed
globule values.

We can more quantitatively compare simulation and experimental studies
of $\alpha$-synuclein by calculating the distributions of inter-residue
distances or, equivalently, the FRET efficiencies.  FRET efficiencies
between residues $i$ and $j$ are obtained from
\begin{equation}
\label{fret}
ET_{\rm eff} = \left\langle \frac{1}{1+\left(\frac{R_{ij}}{R_0}\right)^6}\right\rangle,
\end{equation}
where $R_0 = 54 ~\mbox{Å}$ is the F\"{o}rster distance for the
fluorophore pair in Refs.~\cite{trexler_single_2010,schuler_2002} and
the angle brackets indicate an average over time.  To calculate
$\langle R_{ij} \rangle$ from the FRET efficiencies, one must invert
Eq.~\ref{fret} using the distribution of inter-residue separations
$P(R_{ij})$.  

The FRET efficiencies for the twelve residue pairs from recent smFRET
experiments on $\alpha$-synuclein~\cite{trexler_single_2010} and the
united-atom and coarse-grained simulations are shown in
Fig.~\ref{fig-ETeffs} (a) and (b).  Errors in the inter-residue
separation distributions can occur in both the directly measured
$ET_{\rm eff}$ values and $R_0$. To estimate the errors, we generated
$10$ decoy sets of inter-residue separations using $ET_{\rm eff}$ and
$R_0$ values drawn from distributions accounting for the individual
uncertainties. We then calculated the rms deviation over
each decoy set assuming that we know $R_0$ precisely.
 
We identify several important features in the comparison of the FRET
efficiencies from experiments and simulations in Fig.~\ref{fig-ETeffs}
(a) and (b): 1) The united-atom and coarse-grained models yield
qualitatively similar results for the FRET efficiencies; 2) The FRET
efficiencies for the random walk and pure electrostatics models are
similar to each other and much lower than most of the residue pair
FRET efficiencies from experiments; 3) The FRET efficiencies for the
collapsed globule $\approx 1$ and do not match those from experiments;
and 4) By tuning $\alpha$, we are able to match quantitatively the
FRET efficiencies from the experiments and simulations.

As shown in Fig.~\ref{fig-ETeffs} (c), the rms deviations $\Delta$ between the
FRET efficiencies from the united-atom simulations and smFRET
experiments and between the FRET efficiencies from the coarse-grained
simulations and smFRET experiments are minimized when $\alpha \approx
1.1$.  For the united-atom model, $\alpha \approx 1.1$ gives $\langle
R_g \rangle \approx 33 ~\mbox{Å}$, which is similar to that found
in Ref.~\cite{Nath_conformational_2012}.  The largest deviations in
the FRET efficiencies between the united-atom simulations and smFRET
experiments occur for small inter-residue separations, which are
likely caused by the finite size of the dye molecules.  Note that the
deviations at small inter-residue separations are much weaker for the
coarse-grained simulations.  Thus, we find that it is crucial to
include both electrostatic and attractive hydrophobic interactions in
modeling $\alpha$-synuclein in solution.
 
\begin{figure*}
\noindent \begin{centering}
\includegraphics[width=3in]{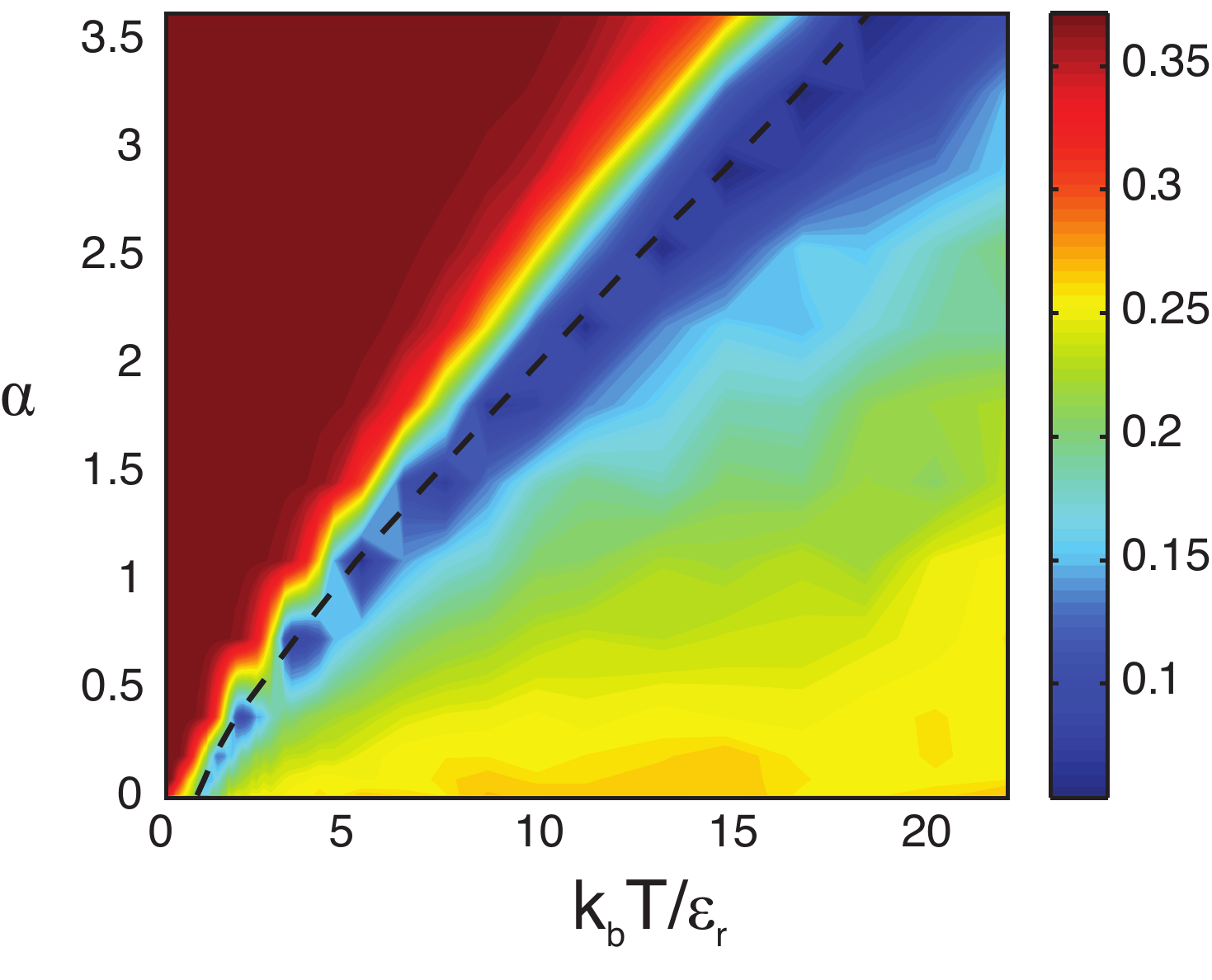}
\par\end{centering}

\caption{\label{cg}RMS deviations $\Delta$ between the coarse-grained
and experimental FRET efficiencies for the twelve residue pairs
considered in Ref.\cite{Nath_conformational_2012} as a function of
$\alpha$ and $k_b T/\epsilon_r$. The dashed line indicates systems
that give $\langle R_g \rangle \simeq 33~\mbox{Å}$.  Note that this
line coincides with the minimum values for the rms deviations.}
\end{figure*}

For the coarse-grained simulations, we also studied the variation of the
FRET efficiencies as a function of temperature (not only at $T=T_0$).  In
Fig.~\ref{cg}, we show the rms deviation between the FRET efficiencies
for the coarse-grained simulations and smFRET experiments for the
twelve residue pairs considered in
Ref.~\cite{Nath_conformational_2012} as a function of $\alpha$ and
$k_b T/\epsilon_r$. We find that the line of $\alpha$ and
$k_bT/\epsilon_r$ values that give $\langle R_g \rangle \simeq
33~\mbox{Å}$ lies in the region where the rms deviations in the FRET
efficiencies are minimized, which indicates that there is a class of 
polymeric structures with similar conformational statistics to that 
of $\alpha$-synuclein.
    
\begin{figure*}
\noindent \begin{centering}
\includegraphics[width=6.5in]{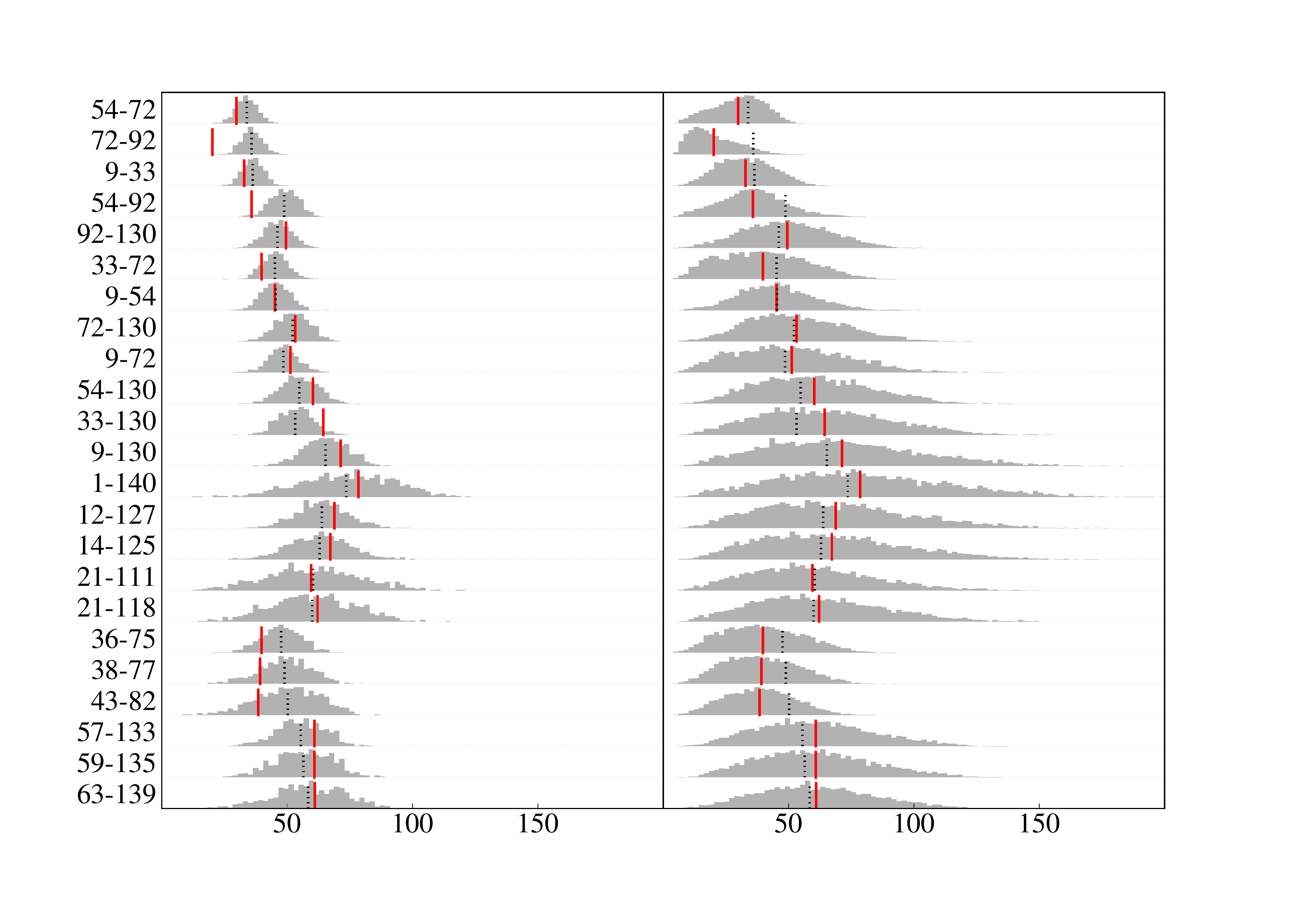}
\par\end{centering}

\caption{\label{rij}Probability distributions for the inter-residue
separations $P(R_{ij})$ for the twelve residue pairs considered in
Ref.~\cite{Nath_conformational_2012} and eleven additional pairs for
experimentally constrained Monte Carlo
(ECMC)~\cite{Nath_conformational_2012} (left) and united-atom 
(with $\alpha=1.1$; right) simulations. The average inter-residue separations
$\langle R_{ij} \rangle$ for the united atom and ECMC simulations are shown
with solid and dashed lines, respectively.}
\end{figure*}

\begin{figure*}
\centering
\begin{subfigure}{3in}
\centering
\includegraphics[width=\textwidth]{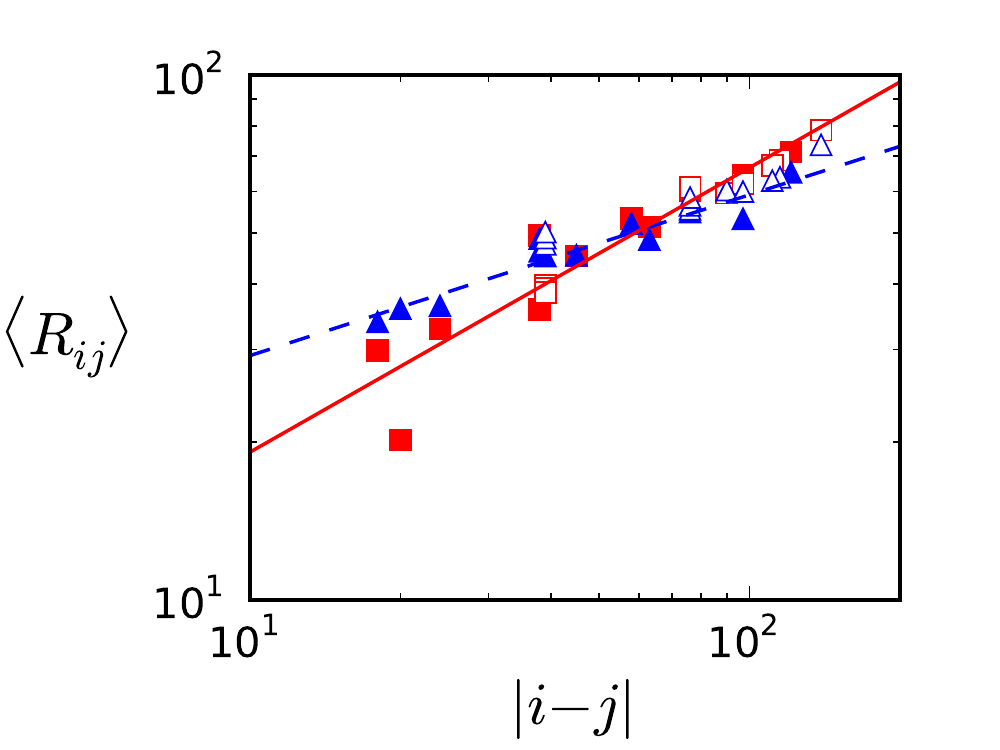}
\end{subfigure}
~
\begin{subfigure}{3in}
\centering
\includegraphics[width=\textwidth]{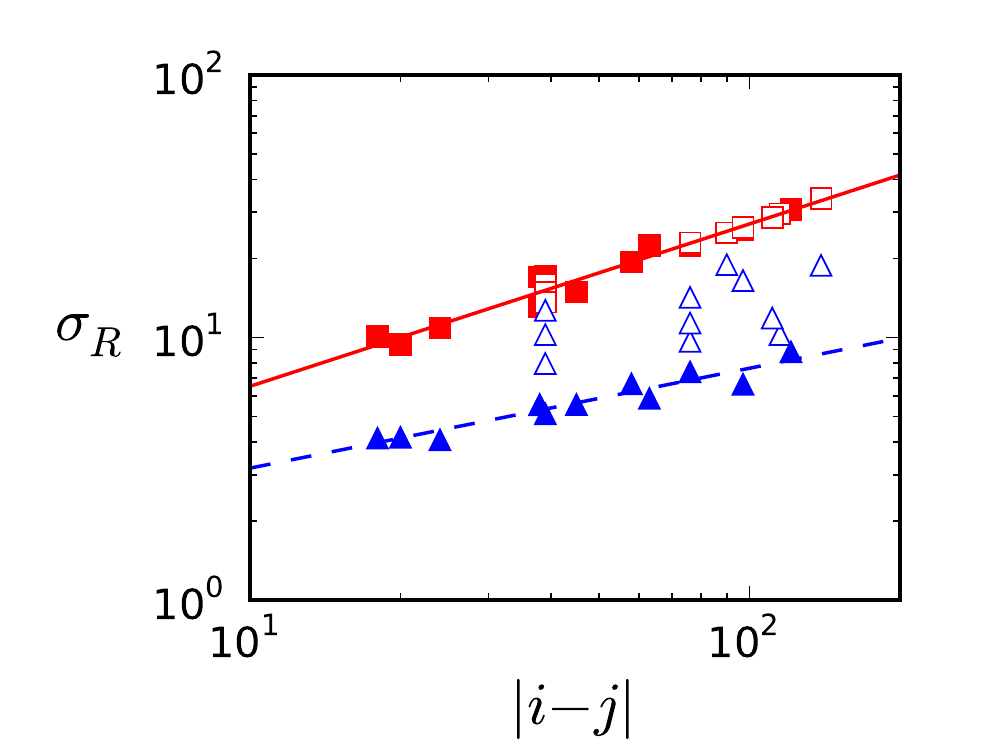}
\end{subfigure}

\caption{\label{fig:8}Average $\langle R_{ij} \rangle$ and standard deviation
$\sigma_R$ of the inter-residue separation distributions in
Fig.~\ref{rij} for the united-atom (squares) and ECMC (triangles)
simulations versus chemical distance between residues $|i-j|$.  The 
filled symbols indicate residue pairs that were considered in 
smFRET experiments~\cite{Nath_conformational_2012} and open 
symbols indicate other pairs. The solid and dashed lines  
have slopes $0.54$ and $0.31$ (left panel) and $0.62$ and $0.38$ (right
panel), respectively.} 
\end{figure*}

In Fig.~\ref{rij}, we compare the inter-residue separation
distributions $P(R_{ij})$ obtained from experimentally constrained
Monte Carlo (ECMC) and united-atom (with $\alpha = 1.1$)
simulations. For the ECMC simulations discussed in detail in
Ref.~\cite{Nath_conformational_2012}, we assumed that $P(R_{ij})$ was
similar to that for a random walk $C_{\alpha}$ model with only
bond-length, bond-angle, and dihedral-angle ($\omega$) constraints and
repulsive Lennard-Jones interactions to obtain $\langle R_{ij}
\rangle$ from the experimentally measured FRET efficiencies.  We find
that $\langle R_{ij} \rangle$ for the ECMC and united-atom simulations
agree to within roughly $10\%$ (Fig.~\ref{fig:8} (left)), however, the
standard deviations differ significantly, as shown in Fig.~\ref{fig:8}
(right).  The standard deviation of $P(R_{ij})$ for the united-atom
simulations is larger than that for the ECMC simulations for all
residue pairs and scales as $\sigma_R \sim |i-j|^{\delta}$ with
$\delta \sim 0.6$ (compared to the excluded volume random walk scaling
exponent $\delta = 0.69$).  Further, $\sigma_R$ for residue pairs that
are not constrained in ECMC do not obey the scaling behavior with
$i-j$ as found for residue pairs that were constrained ($\sigma_R \sim
|i-j|^{\delta}$ with $\delta \sim
0.4$~\cite{Nath_conformational_2012}).
   
\section{Conclusions and Future Directions}
\label{conclusions}

\begin{figure*}
\noindent \begin{centering}
\includegraphics[width=6.5in]{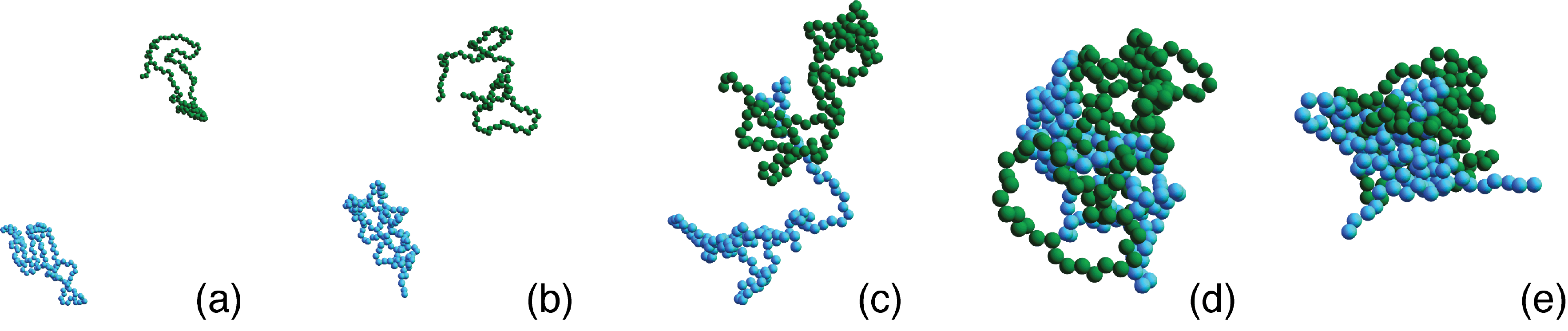}
\par\end{centering}
\caption{\label{aggregate}Snapshots from preliminary aggregation
studies of two monomeric $\alpha$-synuclein proteins (dark green and
light blue) using coarse-grained simulations with the temperature set
so that $\langle R_g \rangle \approx 33 ~\mbox{Å}$ at $\alpha =
1.1$ (for individual protein monomers) for (a) $\alpha=0.7$ (b) $1.1$,
(c) $1.3$, (d) $1.5$, and (e) $1.8$.}
\end{figure*}

We have shown that we are able to accurately model the conformational
dynamics ({\it i.e.} the inter-residue separations) of the IDP
$\alpha$-synuclein at temperature $T_0 = 293 {\rm K}$ and neutral pH
using all-atom, united-atom, and coarse-grained Langevin dynamics
simulations.  Our results show that the structure of
$\alpha$-synuclein is intermediate between that for random walks and
collapsed globules with the rms separation $\sigma_R$ between residues
$i$ and $j$ scaling as $|i-j|^{\delta}$ with $\delta \sim 0.6$.  The
{\it calibrated Langevin dynamics simulations} presented here have the
advantage over constraint methods in that physical forces act on all
residues, not only on residue pairs that are monitored experimentally,
and can be tuned to match FRET efficiencies from experiments.  In
future work, we will employ calibrated Langevin dynamics simulations
to study the conformational dynamics of $\alpha$-synuclein at low pH
and the interaction and association between two or more
$\alpha$-synuclein monomers as a function of pH to identify mechanisms
for $\alpha$-synuclein oligomerization. In preliminary calibrated
coarse-grained Langevin dynamics simulations, we find that two monomeric
$\alpha$-synuclein proteins only associate for sufficiently strong
attractive hydrophobic interactions ($\alpha \ge 1.1$), as shown
in Fig.~\ref{aggregate}.

\section{Acknowledgments}

This research was supported by the National Science Foundation under
Grant Nos. DMR-1006537 (CO, CS), PHY-1019147 (WS), BIO-0919853 (ER), and the
Raymond and Beverly Sackler Institute for Biological, Physical, and
Engineering Sciences (CO, ER).  This work also benefited from the facilities
and staff of the Yale University Faculty of Arts and Sciences High
Performance Computing Center and NSF Grant No. CNS-0821132 that
partially funded acquisition of the computational facilities.

\providecommand{\url}[1]{\texttt{#1}}
\providecommand{\urlprefix}{}
\providecommand{\foreignlanguage}[2]{#2}
\providecommand{\Capitalize}[1]{\uppercase{#1}}
\providecommand{\capitalize}[1]{\expandafter\Capitalize#1}
\providecommand{\bibliographycite}[1]{\cite{#1}}
\providecommand{\bbland}{and}
\providecommand{\bblchap}{chap.}
\providecommand{\bblchapter}{chapter}
\providecommand{\bbletal}{et~al.}
\providecommand{\bbleditors}{editors}
\providecommand{\bbleds}{eds.}
\providecommand{\bbleditor}{editor}
\providecommand{\bbled}{ed.}
\providecommand{\bbledition}{edition}
\providecommand{\bbledn}{ed.}
\providecommand{\bbleidp}{page}
\providecommand{\bbleidpp}{pages}
\providecommand{\bblerratum}{erratum}
\providecommand{\bblin}{in}
\providecommand{\bblmthesis}{Master's thesis}
\providecommand{\bblno}{no.}
\providecommand{\bblnumber}{number}
\providecommand{\bblof}{of}
\providecommand{\bblpage}{page}
\providecommand{\bblpages}{pages}
\providecommand{\bblp}{p}
\providecommand{\bblphdthesis}{Ph.D. thesis}
\providecommand{\bblpp}{pp}
\providecommand{\bbltechrep}{Tech. Rep.}
\providecommand{\bbltechreport}{Technical Report}
\providecommand{\bblvolume}{volume}
\providecommand{\bblvol}{Vol.}
\providecommand{\bbljan}{January}
\providecommand{\bblfeb}{February}
\providecommand{\bblmar}{March}
\providecommand{\bblapr}{April}
\providecommand{\bblmay}{May}
\providecommand{\bbljun}{June}
\providecommand{\bbljul}{July}
\providecommand{\bblaug}{August}
\providecommand{\bblsep}{September}
\providecommand{\bbloct}{October}
\providecommand{\bblnov}{November}
\providecommand{\bbldec}{December}
\providecommand{\bblfirst}{First}
\providecommand{\bblfirsto}{1st}
\providecommand{\bblsecond}{Second}
\providecommand{\bblsecondo}{2nd}
\providecommand{\bblthird}{Third}
\providecommand{\bblthirdo}{3rd}
\providecommand{\bblfourth}{Fourth}
\providecommand{\bblfourtho}{4th}
\providecommand{\bblfifth}{Fifth}
\providecommand{\bblfiftho}{5th}
\providecommand{\bblst}{st}
\providecommand{\bblnd}{nd}
\providecommand{\bblrd}{rd}
\providecommand{\bblth}{th}

\end{multicols}

\appendix

\section{Calibration of Atom Sizes}
\label{apx:atom-sizes}

\begin{figure}[H]
\noindent \begin{centering}
\includegraphics[width=4in]{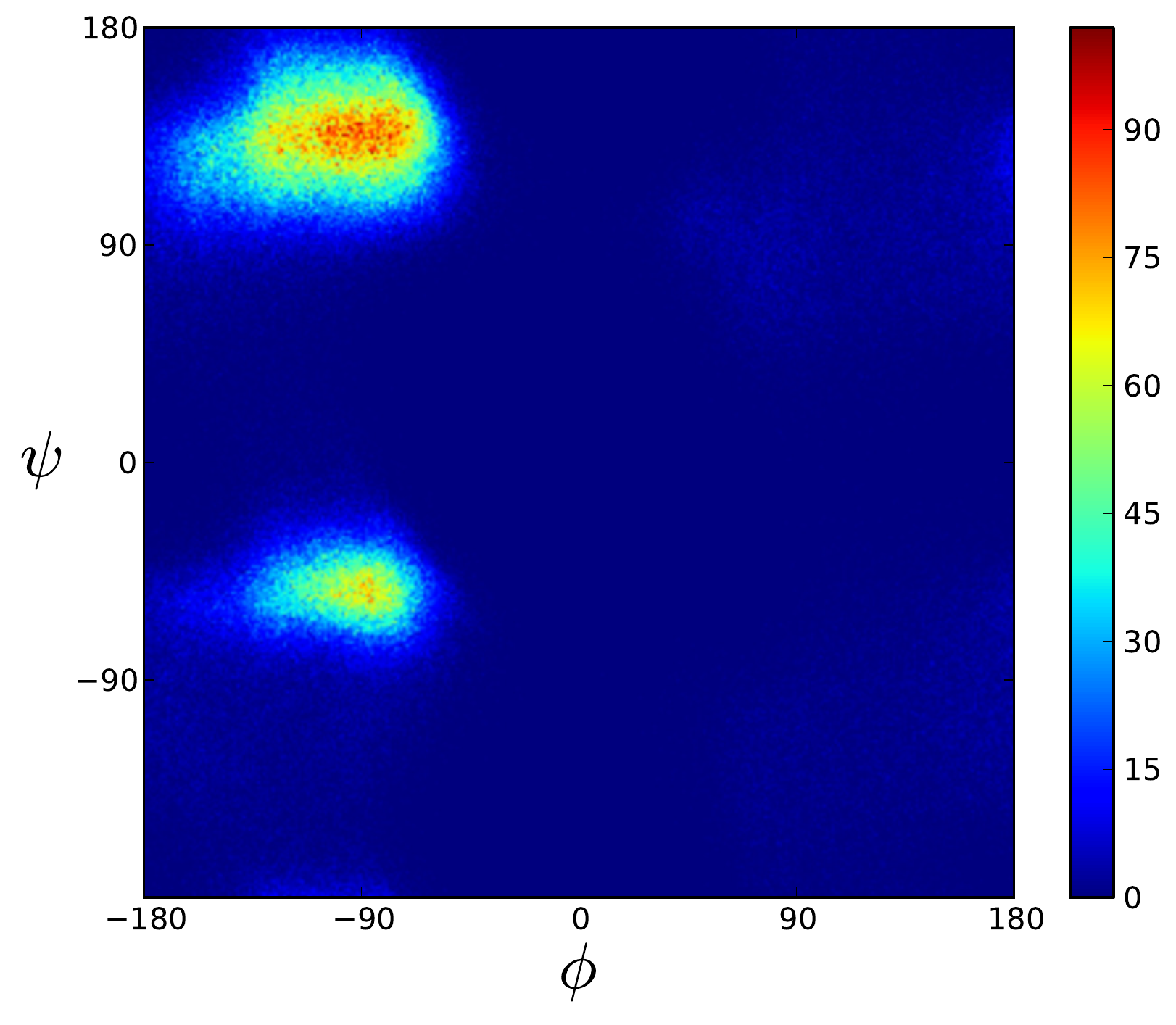}
\par\end{centering}

\caption{\label{fig:Ramachandran-AA}Ramachandran plot for the backbone 
dihedral angles $\phi$ and $\psi$ obtained from the all-atom
random walk simulations with no
attractive hydrophobic and electrostatic interactions, and atom sizes
given in Ref.~\cite{zhou_power_2012}. The highly populated $\phi$ and 
$\psi$ angles indicate $\beta$-sheet (upper left) and $\alpha$-helix
(lower left) conformations.}
\end{figure}

\begin{figure}[H]
\hfill
\begin{subfigure}{3in}
\centering
\includegraphics[width=\textwidth]{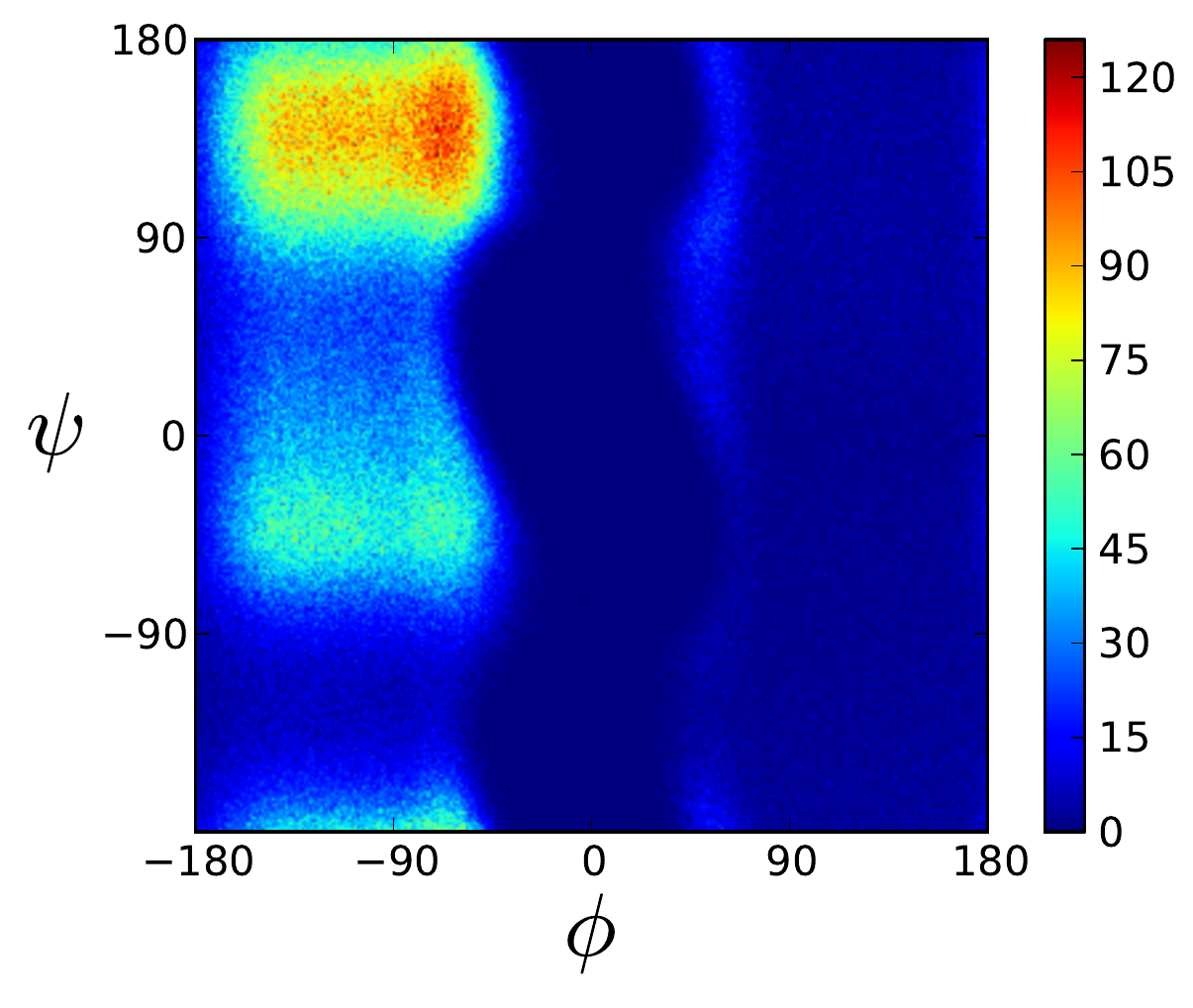}
\end{subfigure}
\hfill
\begin{subfigure}{3in}
\centering
\includegraphics[width=\textwidth]{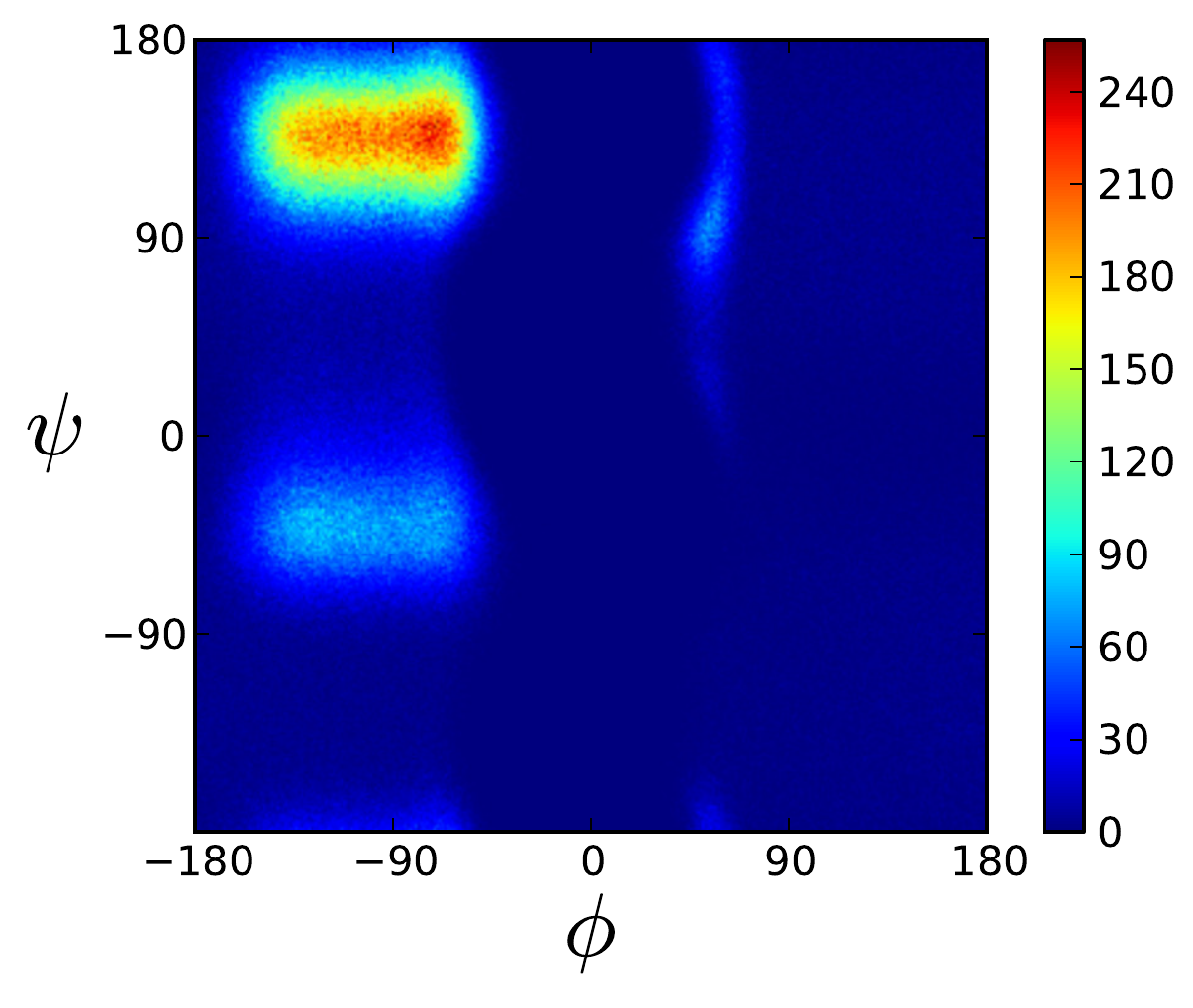}
\end{subfigure}
\hfill

\par

\hfill
\begin{subfigure}{3in}
\centering
\includegraphics[width=\textwidth]{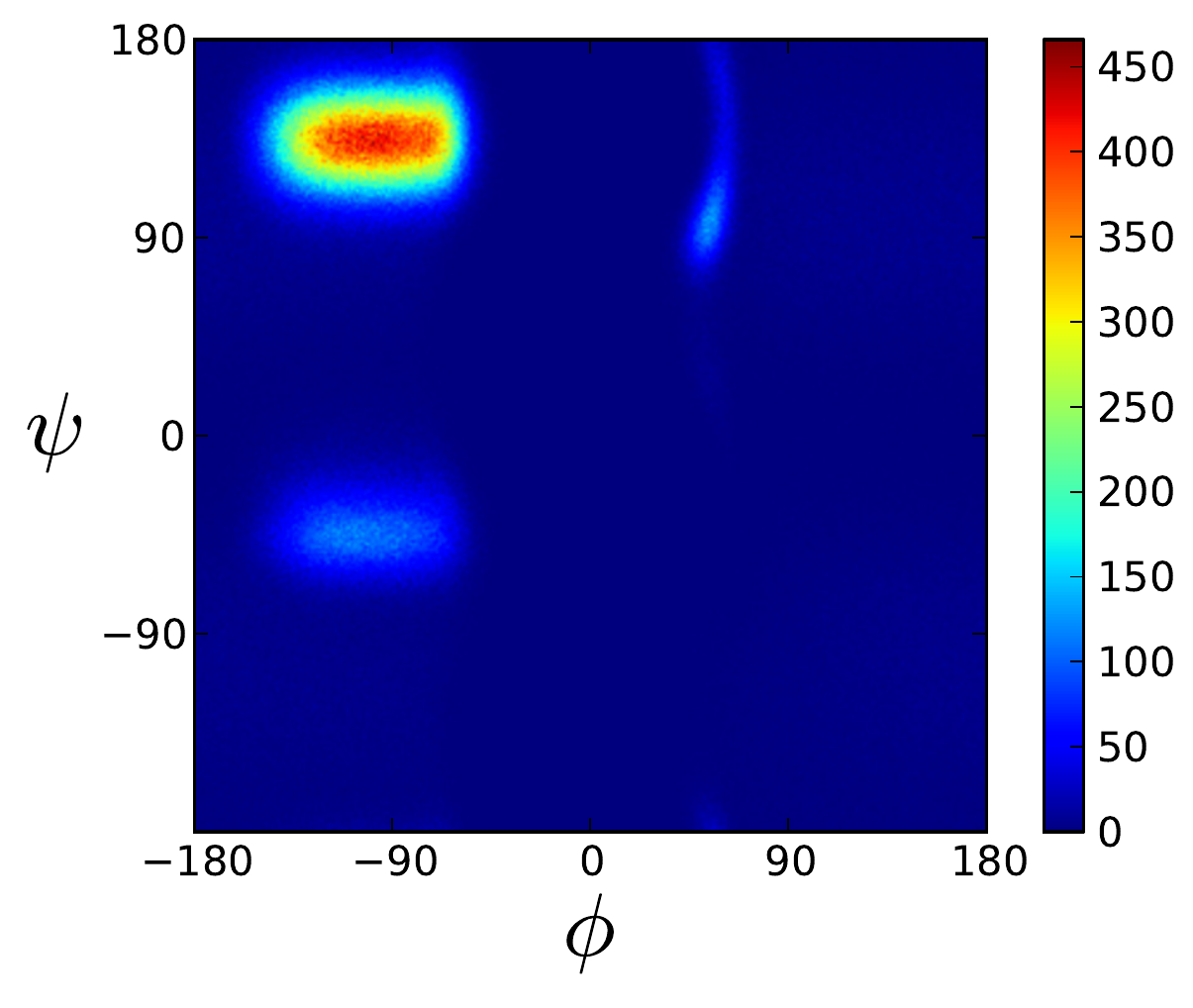}
\end{subfigure}
\hfill
\begin{subfigure}{3in}
\centering
\includegraphics[width=\textwidth]{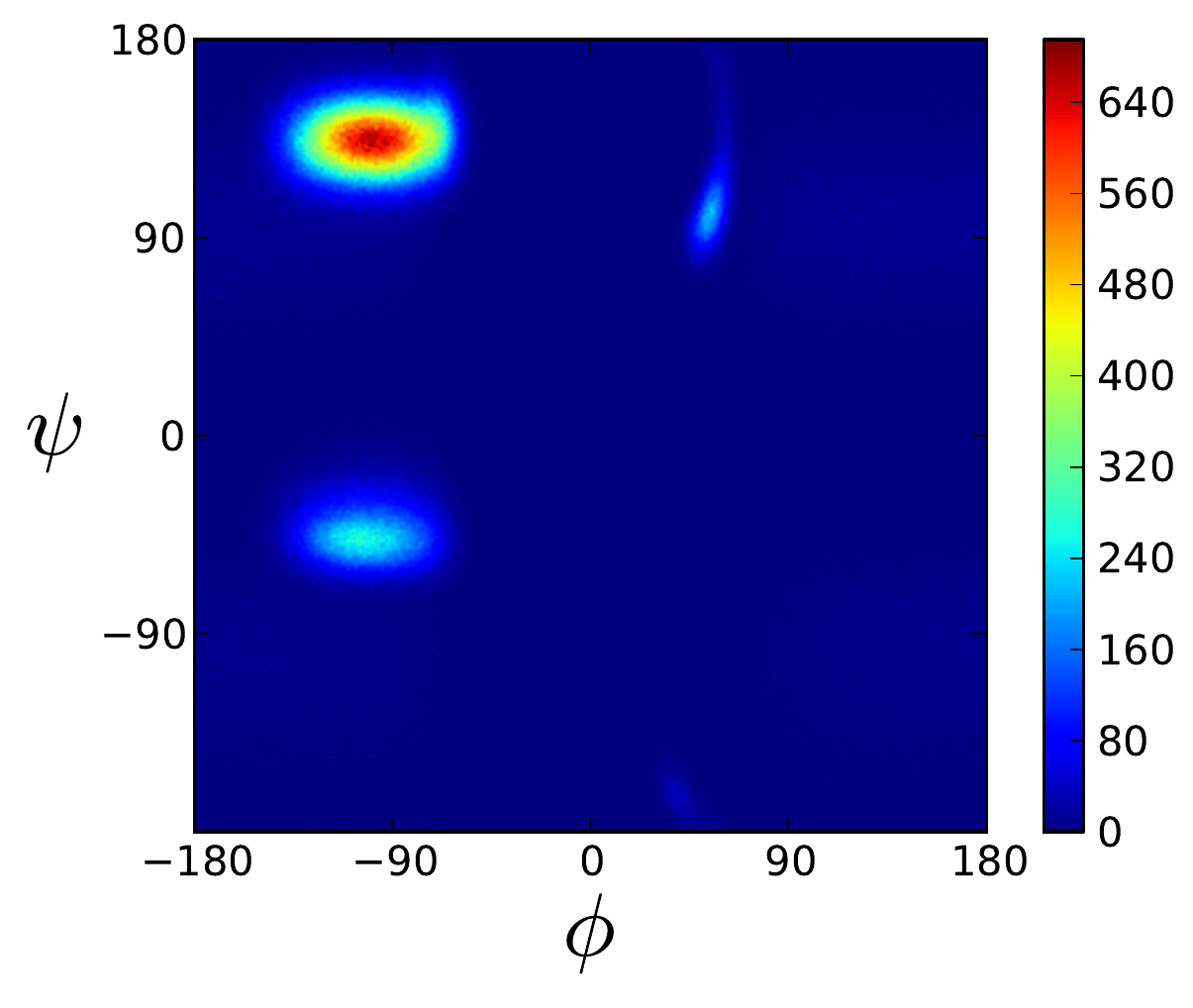}
\end{subfigure}
\hfill

\par
\centering
\begin{subfigure}{3in}
\centering
\includegraphics[width=\textwidth]{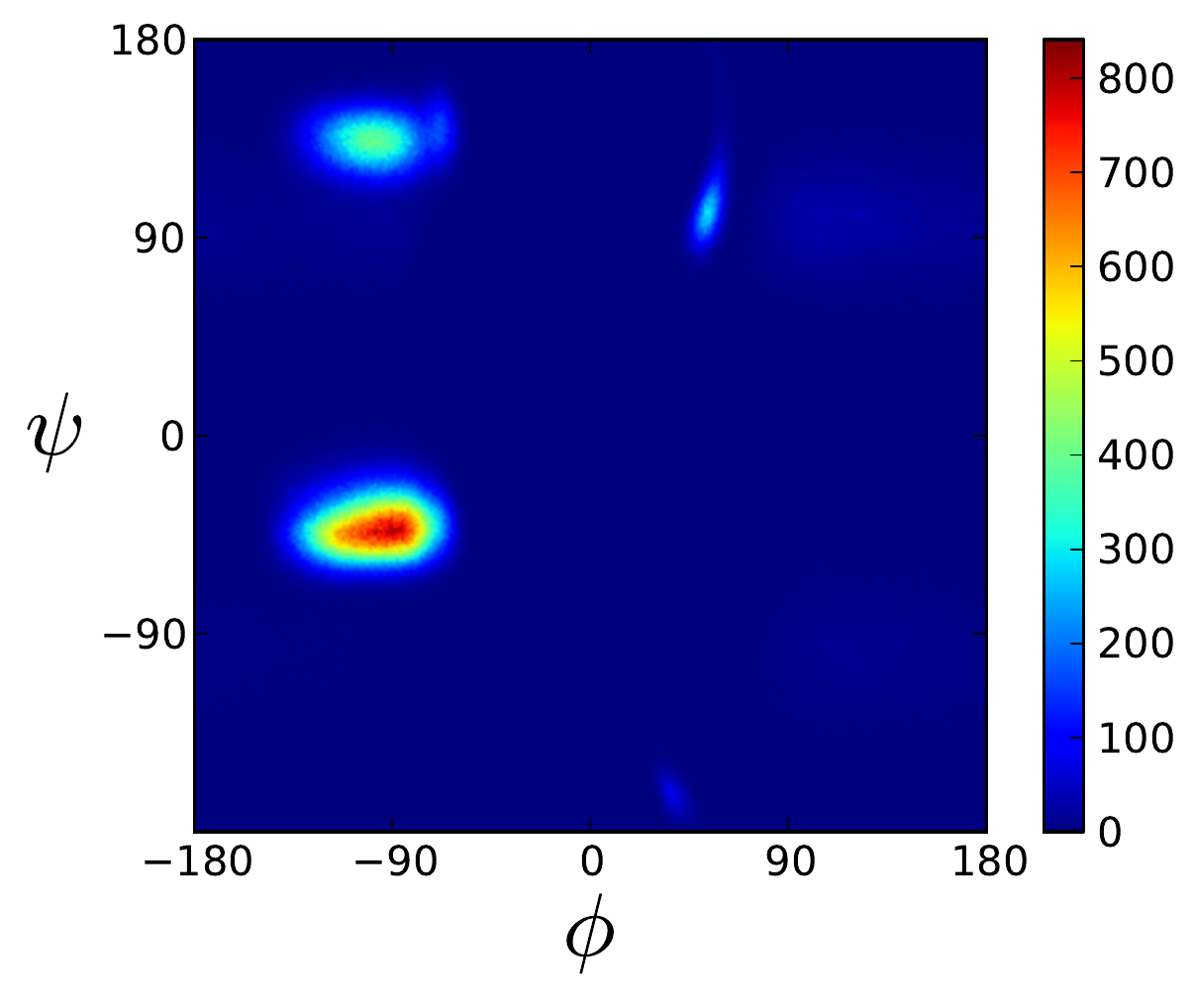}
\end{subfigure}

\caption{\label{fig:Ramachandran} Ramachandran plot for the backbone
dihedral angles $\phi$ and $\psi$ obtained from the united-atom random
walk simulations with no attractive hydrophobic and electrostatic
interactions and atom sizes $0.8$ (upper left), $0.85$ (upper right),
$0.9$ (middle left), $0.95$ (middle right), and $1.0$ (bottom) times
those given in Ref.~\cite{richards_interpretation_1974}.}
\end{figure}

In this Appendix, we test the choice of the atom sizes used in the
all-atom and united-atom models by measuring the Ramachandran
plot~\cite{ramachandran_1963} for the backbone dihedral angles $\phi$
and $\psi$.  In Fig.~\ref{fig:Ramachandran-AA}, we show that the
Ramachandran plot for the random walk all-atom model of
$\alpha$-synuclein with no attractive hydrophobic and electrostatic
interactions and atom sizes from Ref.~\cite{zhou_power_2012} closely
resembles that for dipeptides with highly populated $\alpha$-helix 
and $\beta$-sheet regions.  In Fig.~\ref{fig:Ramachandran}, we
show the Ramachandran plots for the backbone dihedral angles $\phi$
and $\psi$ obtained from the random walk united-atom model of
$\alpha$-synuclein with no attractive hydrophobic and electrostatic
interactions and atom sizes $0.8$, $0.85$, $0.9$, $0.95$, and $1.0$ 
times those from Ref.~\cite{richards_interpretation_1974}.  We find 
that the Ramachandran plot for united-atom model with a factor of $0.9$   
for the atom sizes is similar to that for the all-atom model.   

\section{Robustness of the Hydrophobic Interactions}
\label{apx:robust}

\begin{figure}[H]
\hfill
\begin{subfigure}{3in}
\centering
\includegraphics[width=\textwidth]{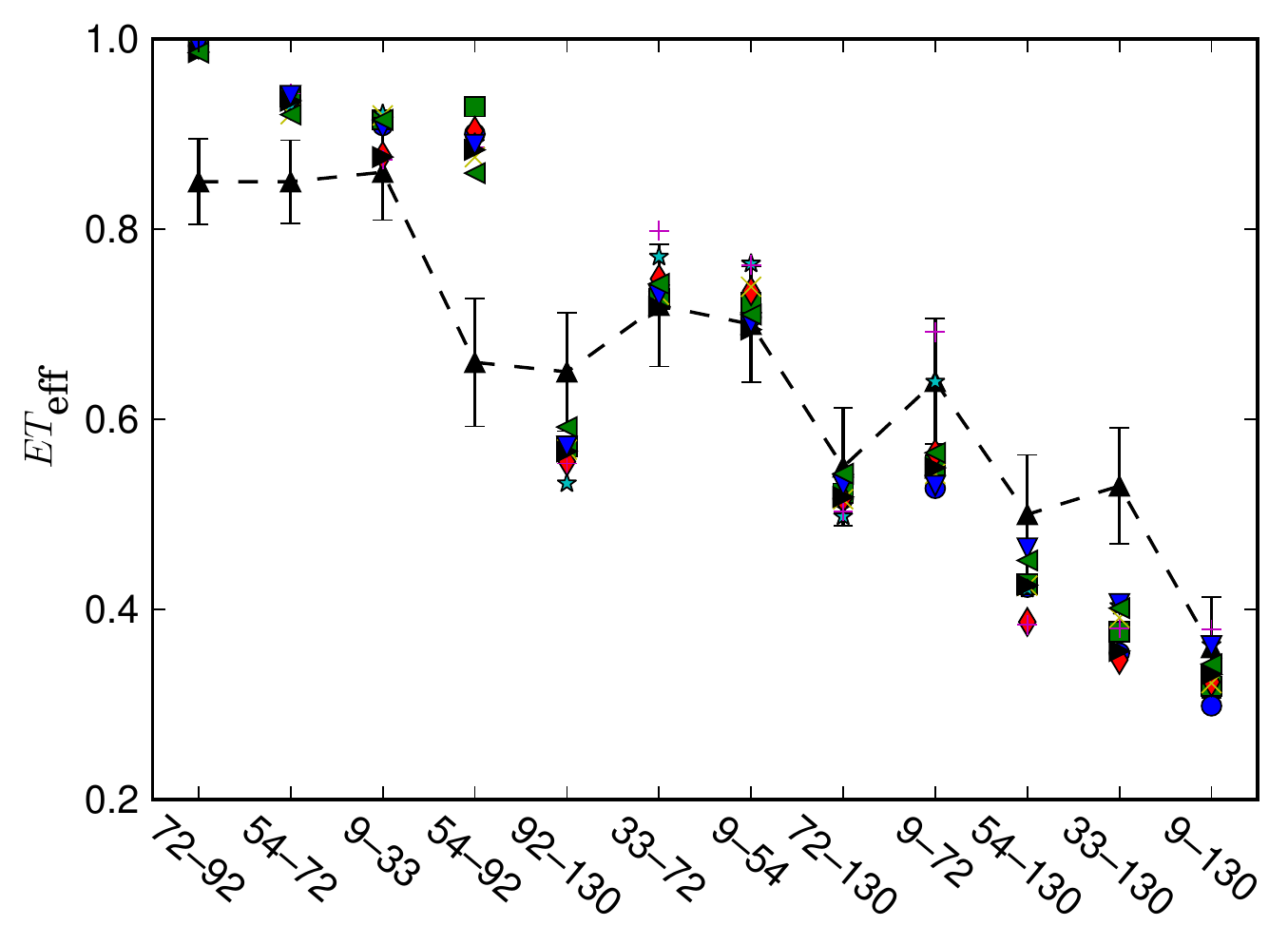}
\end{subfigure}
\hfill
\begin{subfigure}{3in}
\centering
\includegraphics[width=\textwidth]{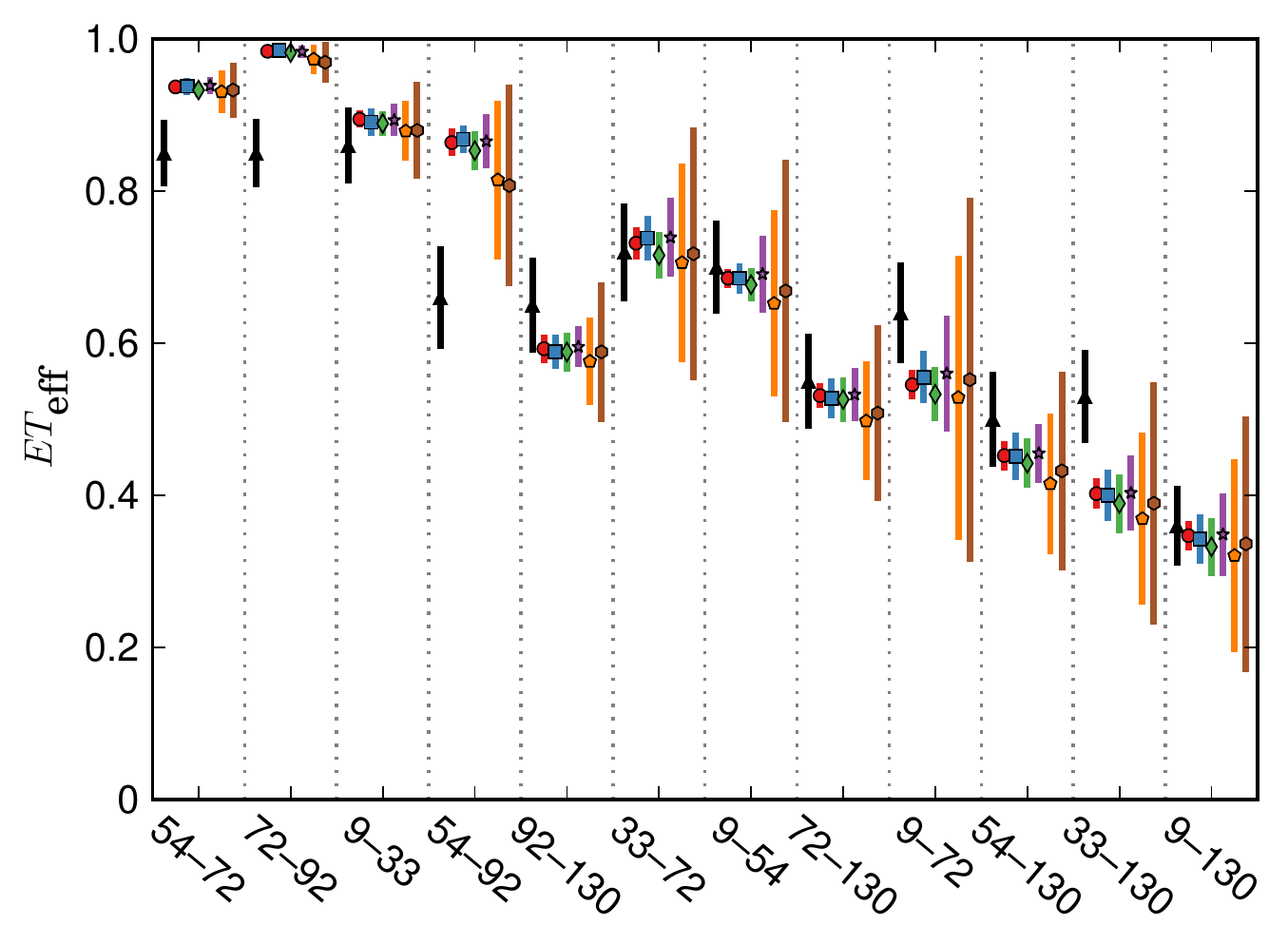}
\end{subfigure}
\hfill

\caption{\label{fig:h_i_robustness} (left) FRET efficiencies $ET_{\rm
eff}$ for the twelve residue pairs considered in
Ref.~\cite{Nath_conformational_2012} from smFRET experiments (upward
triangles) and united-atom simulations with $\alpha$ set so that $R_g
\approx 33~\mbox{Å}$ and $\sigma^a = 4.4~\mbox{Å}$ (circles),
$4.6~\mbox{Å}$ (squares), $4.8~\mbox{Å}$ (diamonds),
$5.0~\mbox{Å}$ (stars), and $5.2~\mbox{Å}$ (pentagons).  (right)
FRET efficiencies $ET_{\rm eff}$ for the twelve residue pairs
considered in Ref.~\cite{Nath_conformational_2012} for the united-atom
simulations for $\alpha = 1.1$ and varying hydrophobicity indices
$h_i' = h_i + \Delta h$, where $\Delta h$ is chosen from a zero-mean
Gaussian distribution with standard deviation $0.0$ (circles), $0.02$
(squares), $0.05$ (diamonds), $0.05$ (diamonds), $0.1$ (stars), $0.3$
(pentagons), $0.5$ (hexagons). The average $ET_{\rm eff}$ and its
standard deviation for $32$ samples are shown for each $\Delta
h$. }
\end{figure}

In this Appendix, we study the sensitivity of the FRET efficiencies
for the united-atom simulations to small variations in the lengthscale
$\sigma^a$ above which the attractive hydrophobic interactions are
nonzero and relative strengths $h_i$ of the attractive hydrophobic
interactions for different residues.  In Fig.~\ref{fig:h_i_robustness}
(left), we show that the FRET efficiencies for the twelve residue
pairs show only small variations with $\sigma^a$ over the range from
$4.3~\mbox{Å}$ to $5.2~\mbox{Å}$ (except for $9$-$72$ with
$\sigma^a=4.3~\mbox{Å}$).  In Fig.~\ref{fig:h_i_robustness}
(right), we show that the FRET efficiencies for the twelve residue
pairs are robust for $\Delta h < 0.5$.  

\begin{thebibliography}{10}

\bibitem{vucetic_flavors_2003}
Vucetic, S.; Brown, C.~J.; Dunker, A.~K.; Obradovic, Z. \emph{Proteins:
  Structure, Function, and Bioinformatics} \textbf{2003}, \emph{52}, 573--584.

\bibitem{sugase_mechanism_2007}
Sugase, K.; Dyson, H.~J.; Wright, P.~E. \emph{Nature} \textbf{2007},
  \emph{447}, 1021--1025.

\bibitem{uversky_intrinsically_2008}
Uversky, V.~N.; Oldfield, C.~J.; Dunker, A.~K. \emph{Annu. Rev. Biophys.}
  \textbf{2008}, \emph{37}, 215--246.

\bibitem{dedmon_mapping_2004}
Dedmon, M.~M.; {Lindorff-Larsen}, K.; Christodoulou, J.; Vendruscolo, M.;
  Dobson, C.~M. \emph{J. Am. Chem. Soc.} \textbf{2004}, \emph{127}, 476--477.

\bibitem{vilar_fold_2008}
Vilar, M.; Chou, H.-T.; Lührs, T.; Maji, S.~K.; Riek-Loher, D.; Verel, R.;
  Manning, G.; Stahlberg, H.; Riek, R. \emph{Proceedings of the National
  Academy of Sciences} \textbf{2008}, \emph{105}, 8637--8642.

\bibitem{eliezer_conformational_2001}
Eliezer, D.; Kutluay, E.; Bussell~Jr, R.; Browne, G. \emph{Journal of Molecular
  Biology} \textbf{2001}, \emph{307}, 1061--1073.

\bibitem{li_conformational_2002}
Li, J.; Uversky, V.~N.; Fink, A.~L. \emph{{NeuroToxicology}} \textbf{2002},
  \emph{23}, 553--567.

\bibitem{tsigelny_dynamics_2007}
Tsigelny, I.~F.; {Bar-On}, P.; Sharikov, Y.; Crews, L.; Hashimoto, M.; Miller,
  M.~A.; Keller, S.~H.; Platoshyn, O.; Yuan, J. X.~J.; Masliah, E. \emph{{FEBS}
  Journal} \textbf{2007}, \emph{274}, 1862--1877.

\bibitem{uversky_effects_2005}
Uversky, V.~N.; Yamin, G.; Munishkina, L.~A.; Karymov, M.~A.; Millett, I.~S.;
  Doniach, S.; Lyubchenko, Y.~L.; Fink, A.~L. \emph{Molecular Brain Research}
  \textbf{2005}, \emph{134}, 84--102.

\bibitem{ullman_explaining_2011}
Ullman, O.; Fisher, C.~K.; Stultz, C.~M. \emph{Journal of the American Chemical
  Society} \textbf{2011}, \emph{133}, 19536--19546.

\bibitem{trexler_single_2010}
Trexler, A.~J.; Rhoades, E. \emph{Biophysical Journal} \textbf{2010},
  \emph{99}, 3048--3055.

\bibitem{Nath_conformational_2012}
Nath, A.; Sammalkorpi, M.; DeWitt, D.~C.; Trexler, A.~J.; S., E.-G.; O'Hern,
  C.~S.; Rhoades, E. \emph{Biophysical Journal} \textbf{2012}, To appear.

\bibitem{dunbrack_jr._bayesian_1997}
Dunbrack~Jr., R.~L.; Cohen, F.~E. \emph{Protein Science} \textbf{1997},
  \emph{6}, 1661--1681.

\bibitem{zhou_power_2012}
Zhou, A.~Q.; {O'Hern}, C.~S.; Regan, L. \emph{Biophysical Journal}
  \textbf{2012}, \emph{102}, 2345--2352.

\bibitem{ramachandran_1963}
Ramachandran, G.; Ramakrishnan, C.; Sasisekharan, V. \emph{Journal of Molecular
  Biology} \textbf{1963}, \emph{7}, 95 -- 99.

\bibitem{monera_relationship_1995}
Monera, O.~D.; Sereda, T.~J.; Zhou, N.~E.; Kay, C.~M.; Hodges, R.~S.
  \emph{Journal of Peptide Science} \textbf{1995}, \emph{1}.

\bibitem{oostenbrink_2004}
Oostenbrink, C.; Villa, A.; Mark, A.~E.; Van~Gunsteren, W.~F. \emph{Journal of
  Computational Chemistry} \textbf{2004}, \emph{25}, 1656--1676.

\bibitem{richards_interpretation_1974}
Richards, F. \emph{Journal of Molecular Biology} \textbf{1974}, \emph{82},
  1--14.

\bibitem{ermak}
Ermak, D.~L.; Buckholz, H. \emph{Journal of Computational Physics}
  \textbf{1980}, \emph{35}, 169 -- 182.

\bibitem{ulmer_structure_2005}
Ulmer, T.~S.; Bax, A.; Cole, N.~B.; Nussbaum, R.~L. \emph{Journal of Biological
  Chemistry} \textbf{2005}, \emph{280}, 9595--9603.

\bibitem{uversky_trimethylamine-n-oxide-induced_2001}
Uversky, V.~N.; Li, J.; Fink, A.~L. \emph{{FEBS} Letters} \textbf{2001},
  \emph{509}, 31--35.

\bibitem{tashiro_characterization_2008}
Tashiro, M.; Kojima, M.; Kihara, H.; Kasai, K.; Kamiyoshihara, T.; Uéda, K.;
  Shimotakahara, S. \emph{Biochemical and Biophysical Research Communications}
  \textbf{2008}, \emph{369}, 910--914.

\bibitem{rekas_structure_2010}
Rekas, A.; Knott, R.; Sokolova, A.; Barnham, K.; Perez, K.; Masters, C.; Drew,
  S.; Cappai, R.; Curtain, C.; Pham, C. \emph{European Biophysics Journal}
  \textbf{2010}, \emph{39}, 1407--1419.

\bibitem{salmon_nmr_2010}
Salmon, L.; Nodet, G.; Ozenne, V.; Yin, G.; Jensen, M.~R.; Zweckstetter, M.;
  Blackledge, M. \emph{Journal of the American Chemical Society} \textbf{2010},
  \emph{132}, 8407--8418.

\bibitem{schuler_2002}
Schuler, B.; Lipman, E.~A.; Eaton, W.~A. \emph{Nature} \textbf{2002},
  \emph{419}, 743--747.

\end{thebibliography}
\end{document}